\documentclass[preprint]{aastex}

\newcommand{\Htwo}{H$_2$}
\newcommand{\FUSE}{{\it FUSE}}
\newcommand{\FUSElong}{{\it Far Ultraviolet Spectroscopic
       Explorer}}

\newcommand{\kms}{km~s{$^{-1}$}}

\newcommand\etal{et~al.}
\newcommand{\cd}{cm$^{-2}$}
\newcommand{\W}{$W_{\lambda}$}

\newcommand{\NHtwo}{N$_{\rm H2}$}
\newcommand{\NHI}{N$_{\rm HI}$}

\begin{document}

\title{A \FUSE\ SURVEY OF INTERSTELLAR MOLECULAR HYDROGEN
TOWARD HIGH-LATITUDE AGN}

\author{
Kristen Gillmon\altaffilmark{1,2},
J. Michael Shull\altaffilmark{1},
Jason Tumlinson\altaffilmark{3},
\& Charles Danforth\altaffilmark{1}
}

\email{
kristen.gillmon@colorado.edu,
mshull@casa.colorado.edu,
tumlinso@oddjob.uchicago.edu,
charles.danforth@colorado.edu
}

\altaffiltext{1}{Center for Astrophysics and Space Astronomy, Dept.\ of 
   Astrophysical \& Planetary Sciences,
   University of Colorado, 389 UCB, Boulder, CO 80309}
\altaffiltext{2}{Now at Dept.\ of Astronomy,
   University of California, Berkeley, CA 94720}
\altaffiltext{3}{Department of Astronomy \& Astrophysics,
  University of Chicago, 5640 S. Ellis Ave., Chicago, IL 60637}



\begin{abstract}\label{abstract}

We report results from a \FUSElong\ (\FUSE) survey of interstellar
molecular hydrogen (\Htwo) along 45 sight lines to AGN at high 
Galactic latitudes ($|b| > 20^{\circ}$).  Most (39 of 45) of the 
sight lines show detectable Galactic \Htwo\ absorption from
Lyman and Werner bands between 1000 and 1126 \AA, with column 
densities ranging from \NHtwo\ $=10^{14.17-19.82}$~\cd.
In the northern Galactic hemisphere, we identify many regions of
low \NHtwo\ ($\leq 10^{15}$~\cd) between
$\ell =60^{\circ}-180^{\circ}$
and at $b>54^{\circ}$. These ``\Htwo\ holes" provide valuable,
uncontaminated sight lines for extragalactic UV spectroscopy, and
a few may be related to the ``Northern Chimney" (low \ion{Na}{1}
absorption) and  ``Lockman Hole" (low \NHI). A comparison
of high-latitude \Htwo\ with 139 OB-star sight lines surveyed in
the Galactic disk suggests that high-latitude and disk \Htwo\ clouds may 
have different rates of heating, cooling, and UV excitation.  For
rotational states $J = 0$ and 1, the mean excitation temperature
at high latitude, $\langle T_{01}^{\rm hl} \rangle = 124 \pm 8$~K,
is somewhat higher than in the Galactic disk,
$\langle T_{01}^{\rm disk} \rangle= 86 \pm 20$~K.
For $J \geq 2$, the mean $\langle T_{\rm exc} \rangle = 498 \pm 28$~K,
and the column-density ratios, N(3)/N(1), N(4)/N(0), and N(4)/N(2),
indicate a comparable degree of UV excitation in the disk and low halo
for sight lines with \NHtwo\ $\geq 10^{18}$ \cd.  The distribution of
molecular fractions at high latitude shows a transition 
at lower total hydrogen column density
(log~N$_{\rm H}^{\rm hl} \approx 20.38 \pm 0.13$) than in
the Galactic disk (log~N$_{\rm H}^{\rm disk} \approx 20.7$).  If the
UV radiation fields are similar in disk and low halo, this suggests
an enhanced \Htwo\ (dust-catalyzed) formation rate in higher-density,
compressed clouds, which could be detectable as high-latitude, sheetlike
infrared cirrus.

\end{abstract}


\section{INTRODUCTION}\label{intro}

Molecular hydrogen (\Htwo) is the most abundant molecule
in the universe, comprising the majority of mass in
interstellar molecular clouds that eventually form stars.
In the diffuse interstellar medium, with visual
extinctions $A_V \leq 1$~mag, the absorbing clouds exhibit
detectable \Htwo\ lines with molecular fractions, $f_{\rm H2}$,
ranging from $10^{-6}$ at low H~I column density up to
$\sim$40\% (Savage \etal\ 1977; Spitzer \& Jenkins 1975;
Shull \etal\ 2005).  In sight lines with
even greater  extinction, the so-called translucent
sight lines with $A_V = 1-5$ mag (van~Dishoeck \& Black 1986),
the molecular fraction can be as high as 70\% (Rachford \etal\
2002). Even though \Htwo\ plays an important role in
interstellar chemistry, many questions remain about its
distribution, formation, and destruction in protostellar
clouds.  With the 1999 launch of the \FUSElong\ (\FUSE)
satellite, astronomers regained access to far ultraviolet (FUV)
wavelengths below 1126 \AA, needed to study \Htwo\ via
resonance absorption lines from its ground electronic state,
X~$^{1}\Sigma^{+}_{g}$, to the excited states,
B~$^{1}\Sigma_{u}^{+}$ (Lyman bands) and
C~$^{1}\Pi_u$ (Werner bands).  In absorption by the
cold clouds, the observed transitions originate in the
ground ($v = 0$) vibrational state from a range of
rotational states, primarily $J=0$ (para-\Htwo\ with
total nuclear spin $S=0$) and $J=1$ (ortho-\Htwo\ with
nuclear spin $S=1$).  Weaker absorption lines from excited
states, $J \geq 2$, are also detected, which are produced by UV
fluorescence and cascade following photoabsorption in the
Lyman and Werner bands;  see reviews by Spitzer \& Jenkins
(1975) and Shull \& Beckwith (1982).

From the \FUSE\ mission inception, \Htwo\ studies have been an
integral part of the science plan.  The \FUSE\ satellite,
its mission, and its on-orbit performance are described
in Moos \etal\ (2000) and Sahnow \etal\ (2000).
The initial \FUSE\ studies of \Htwo\ appeared among the
Early Release Observations (Shull \etal\ 2000; Snow
\etal\ 2000).  More recent observations include \FUSE\
surveys of \Htwo\ abundance in translucent clouds (Rachford
\etal\ 2002) and diffuse clouds (Shull \etal\ 2005)
throughout the Milky Way disk, in interstellar gas toward
selected quasars (Sembach \etal\ 2001a, 2004), in intermediate
velocity clouds (IVCs) and high velocity clouds (HVCs) 
(Sembach \etal\ 2001b; Richter \etal\ 2001, 2003);
in the Monoceros Loop supernova remnant (Welsh, Rachford,
\& Tumlinson 2002), and in gas throughout the Large and Small
Magellanic Clouds (Tumlinson \etal\ 2002).

Sightlines toward active galactic nuclei (AGN) through the
Galactic halo provide special probes
of diffuse interstellar gas, including large-scale gaseous
structures, IVCs, and HVCs. In contrast, surveys
of OB stars in the Galactic disk typically have much higher
column densities, \NHtwo\ $ > 10^{18}$~\cd, except toward
nearby stars  too bright to be observable with the \FUSE\
detectors.  Therefore, the AGN halo survey is able to probe
fundamental properties of \Htwo\ physics, tied to observable
quantities such as metal abundances, H~I, and IR cirrus.  The
halo survey also probes \Htwo\ in a new, optically thin regime
(\NHtwo\ $<10^{17}$~\cd) inaccessible to \FUSE\ in the Galactic
disk.  These low-column clouds can provide important tests
of models of \Htwo\ formation, destruction, and excitation
in optically-thin environments where FUV radiation dominates
the $J$-level populations.
The targets in this survey were typically not observed for
the primary purpose of detecting \Htwo.  Rather, these
high-latitude AGN were observed to study Galactic
O~VI, for the Galactic D/H survey, as probes of the
interstellar medium (ISM) and intergalactic medium (IGM),
and for their intrinsic interest in AGN studies. The relatively
smooth AGN continua allow us to obtain clean measurements of
interstellar absorption.

Molecular hydrogen often contaminates the UV spectra of
extragalactic sources, even when they are used for other purposes,
such as studies of intervening interstellar and intergalactic gas.
Because \Htwo\ has such a rich band structure between 912--1126~\AA,
one must remove \Htwo\ Lyman and Werner band absorption by using a
model for the column densities, N($J$), in \Htwo\ rotational states
$J = 0-5$ and occasionally even higher.
This \Htwo\ spectral contamination becomes detectable at
log~\NHtwo\ $\geq 14$. It becomes serious at log~\NHtwo\ $\geq 16$,
when the resonance lines from $J = 0$ and $J = 1$ become strong
and selected lines from higher rotational states blend with lines
from the ISM and IGM.  A few \Htwo\ absorption lines blend with key
diagnostic lines (O~VI $\lambda\lambda 1031.926, 1037.617$) and
occasionally with redshifted IGM absorbers.  Our AGN sample is
useful for another reason: we are able to assess the range of
\Htwo\ column densities and corrections to FUV spectra taken to
study the ISM and IGM.  This subtraction was performed in studies
of intergalactic O~VI (Savage \etal\ 2002; Danforth \&
Shull 2005), of high-latitude Galactic O~VI (Wakker \etal\ 2003;
Sembach \etal\ 2003; Collins, Shull, \& Giroux 2004), and of
D/H in high velocity cloud Complex C (Sembach \etal\ 2004).

Observations of the hydrogen molecule, its abundance
fraction in diffuse clouds, and its rotational
excitation provide important physical diagnostics of
diffuse interstellar gas (Shull \& Beckwith 1982).
For example, the rotational temperature, $T_{01}$,
of the lowest two rotational states, $J = 0$ and $J = 1$,
is an approximate measure of the gas kinetic temperature
in cases where collisions with H$^+$ or H$^{\circ}$ are
sufficiently rapid to mix the two states.
In diffuse clouds, the \Htwo\ rotational temperature,
$\langle T_{01} \rangle = 77 \pm 17$~K, measured by
{\it Copernicus} (Savage \etal\ 1977) provides one of the
best indicators of the heating and cooling processes in
the diffuse ISM within 1 kpc of the Sun.  Our \FUSE\ survey
(Shull \etal\ 2005) extends this study to more distant OB
stars in the Galactic disk and finds similar temperatures
$\langle T_{01} \rangle = 86 \pm 20$~K.

For higher rotational states, the level populations are
characterized by an excitation temperature, $T_{\rm exc}$,
that reflects the FUV radiation field between 912--1126~\AA,
which excites and photodissociates \Htwo.  Populations of
the $J \geq 2$ rotational levels in the ground vibrational
and ground electronic state are governed by absorption
in the \Htwo\ Lyman and Werner bands.  These electronic
excitations are followed by UV fluorescence
to the ground electronic state and infrared radiative cascade
through excited vibrational and rotational states,
often interrupted by rotational de-excitation by
H$^{\circ}$--H$_2$ collisions.  In approximately 11\%
of the line photoabsorptions, for a typical ultraviolet
continuum, \Htwo\ decays to the vibrational continuum
and is dissociated.  This peculiarity of \Htwo\ allows
the molecule to ``self-shield" against dissociating FUV
radiation, as the Lyman and Werner absorption lines become
optically thick.  Detailed models of these processes (Black
\& Dalgarno 1973, 1976; Browning, Tumlinson, \& Shull 2003)
can be used to estimate the FUV radiation field and gas density.

In this paper, we present a survey of \Htwo\ toward 45
AGN at Galactic latitudes $|b| > 20^{\circ}$ (see Figure 1).
To produce detectable \Htwo, with $f_{\rm H2} > 10^{-5}$,
these clouds typically need hydrogen gas densities
$n_H>$ 10 cm$^{-3}$, so that the \Htwo\ formation rates are
sufficient to offset UV photodissociation.  At column densities
\NHtwo\ $>10^{15}$~\cd, the molecules begin to self-shield
from UV photodestruction.  The selection of high-latitude AGN
background targets allows us to minimize the amount of \Htwo\ from
the Galactic disk, although we cannot exclude it altogether.  Our
high-latitude survey accompanies a larger survey
(Shull et al.\ 2005) of \Htwo\ toward 139 OB-type stars in the
Galactic disk.  These OB stars at $|b| \leq 10^{\circ}$ typically
show larger \Htwo\ column densities and greater molecular
abundances than the AGN sight lines.  Consequently,
the AGN sight lines provide a better sample of
diffuse \Htwo-bearing clouds above the disk, and perhaps into
the low halo.  Various properties of the gas, including molecular 
fraction, rotational temperature, and connections with infrared 
cirrus clouds (Gillmon \& Shull 2005), suggest that the 
high-latitude sight lines sample a different population of diffuse 
absorbers than the disk survey.  The halo \Htwo\ survey may also 
be representative of sight lines that pass outside the inner disks 
of external galaxies.

The organization of this paper begins with a description
of the AGN sample and our data acquisition, reduction, and
analysis (\S~2).  This is followed in \S~3 by discussion of
the results on \Htwo\ abundances, molecular fractions, and
rotational excitation.  We examine both the rotational
temperature, $T_{01}$, for $J=0$ and $J=1$, and the excitation 
temperature, $T_{\rm exc}$, for higher rotational levels.
We also show the distribution of \Htwo\ in column density
and the spatial distribution throughout the northern Galactic
hemisphere, particularly the conspicuous ``H$_2$ holes" with
\NHtwo\ $\leq 10^{15}$~\cd\ at $b > 54^{\circ}$.
We conclude in \S~4 with a comparison of these high-latitude  
results to those from Galactic disk surveys by {\it Copernicus}
(Savage \etal\ 1977) and \FUSE\ (Shull \etal\ 2005).


\section{DATA ACQUISITION AND REDUCTION}\label{data}

\subsection{\FUSE\ Observations}

Table 1 lists our target names and their Galactic coordinates,
source type, B magnitude, color excess E(B-V), and observational
parameters.  We chose our targets based on their signal-to-noise
ratio (S/N), source type, and Galactic latitude.
All 45 of our targets are at $|b| > 20^{\circ}$,
a subset of the 219 targets selected in Wakker \etal\ (2003)
as candidates for the analysis of Galactic O~VI.
Of the 219 targets, Wakker \etal\ found 102 appropriate
for O~VI analysis, for which they binned the data by 5--15
pixels (the \FUSE\ resolution element is 8--10 pixels or
about 20 \kms) in order to meet an imposed requirement of
$\rm{(S/N)_{bin}>4}$. We set a similar
S/N requirement for our \Htwo\ analysis.  However, we
exclude those targets that required binning by more
than 8 pixels, because the loss of resolution caused by
overbinning becomes problematic for the analysis of
relatively narrow \Htwo\ lines, in comparison with the
typically broad ($\geq 50$ \kms) O~VI profiles.
Of the 219 targets, 55 met our requirement of
$\rm{(S/N)_{bin}>4}$ or $\rm{(S/N)_{pix}>2}$ with
4-pixel binning.  We excluded halo stars and
restrict our sample targets to AGN
plus a few starburst galaxies in order to minimize the
difficulties presented by stellar continua. We checked
all targets that were excluded for low S/N to see whether
additional \FUSE\ data  had been acquired since the Wakker
\etal\ (2003) survey. The additional data increased
the S/N for seven of these targets sufficiently to include
them in our sample.

All observations were obtained in time-tag (TTAG) mode,
using the $30'' \times 30''$ LWRS aperture.  The
resolution of \FUSE\ varies from $R = \lambda / \Delta \lambda
= 15,000 - 20,000$ across the far-UV band, and it can also
vary between observations.  However, our results rely on
measured equivalent widths that are effectively independent
of instrumental resolution.  Data were retrieved from the 
archive and reduced locally using CalFUSE v2.4.  The CalFUSE 
processing is described in the FUSE Observers Guide 
(http://fuse.pha.jhu.edu).  Raw exposures within a single 
\FUSE\ observation were coadded by channel mid-way through the 
pipeline.  This can produce a significant improvement in data
quality for faint sources (such as AGN) since the combined 
pixel file has higher S/N than the individual exposures, and 
consequently the extraction apertures are more likely to be
placed correctly.  Combining exposures also speeds up the 
data reduction time substantially.  

Because \FUSE\ has no internal wavelength calibration source,
all data are calibrated using a wavelength solution derived from 
in-orbit observations of sources with well-studied interstellar
components (Sahnow \etal\ 2000).  This process leads to relative
wavelength errors of up to 20 \kms\ across the band, in addition 
to a wavelength zero point that varies between observations.   
For targets with data from multiple 
observations, the spectra from separate observations were coaligned and 
coadded to generate a final spectrum for data channels
LiF1a and LiF2a (the other six data channels were not used in 
our analysis).  The coalignment was performed on one single 
line per channel: \ion{Ar}{1} at 1048.22 \AA\ for LiF1a, and 
Lyman (1-0) P(3) at 1099.79 \AA\ for LiF2a.  Since we did not correct 
for possible relative wavelength errors across the wavelength 
range before coadding, our method of using a single line to perform 
the alignment may blur out weak lines at wavelengths well away from 
1048.22 \AA\ (LiF1a) and 1099.79 \AA\ (LiF2a).

Once the final coadded spectra have been obtained, there may still be
errors in the wavelength solution.
Because the 21~cm spectra are acquired with ground-based telescopes, 
their wavelength solution is more accurate than that of \FUSE\ data.  
We attempted to correct the \FUSE\ wavelength solution by 
plotting the 21~cm emission line and \ion{Ar}{1} 1048.22 \AA\ absorption 
line in velocity space and shifting the \FUSE\ spectra to align the 
centroids.  Although it is not necessarily the case that the neutral
gas (H~I) must align in velocity with low-ionization metals (Ar~I, Si~II, 
Fe~II) or with \Htwo, it is a plausible assumption that provides a 
standard technique for assigning the \FUSE\ wavelength scale. Although 
relative wavelength errors may exist across the full wavelength range,  
we place our trust in the wavelength solution near 1048 \AA\ and 
use \Htwo\ lines in the (4-0) and (5-0) Lyman bands between 
1036--1062 \AA\ to determine the velocity of the \Htwo\ clouds (see \S~2.2).  
The S/N of the co-added data ranges from 2--11 per pixel, and it also varies 
with spectral resolution, which is not fixed in our survey. Most of the 
data were binned by 4 pixels before analysis, with the rare case of 
binning by 2 or 8 pixels.

\subsection{Data Analysis}

In our analysis of \Htwo, we use the signature absorption bands in the 
\FUSE\ LiF channels between 1000 and 1126 \AA.  These bands arise from 
the Lyman and Werner electronic transitions from the ground ($v=0$) 
vibrational state and rotational states $J = 0-7$.  Wavelengths, 
oscillator strengths ($f$), and damping constants for these lines were
taken from Abgrall \etal\ (1993a,b).  Because of the rapid decrease in 
\FUSE\ sensitivity and decreased S/N in the SiC channels at wavelengths 
less than 1000 \AA, we restrict our analysis to ten vibrational-rotational 
bands, Lyman (0-0) to (8-0) and Werner (0-0).  Additional information
can be gleaned from higher vibrational bands ($\lambda < 1000$~\AA) in 
the SiC channels, but we do not present this data here.  
To maximize efficiency, we restricted our analysis to the LiF1a
and LiF2a channels, which provide the best combination
of wavelength coverage and sensitivity.  These channels provide
higher S/N than the SiC channels in this wavelength range and
include the Lyman (2-0) and (1-0) bands.

Seven AGN sight lines show \Htwo\ absorption in $J=4$, and
absorption lines from $J=5$ were detected toward
one target (HS~0624+6907).  Approximately one-third (15)
of the sight lines show substantial \Htwo\ column densities,
ranging from log~\NHtwo\ = 18.0 -- 19.8,
with damping wings in the $J=0$ and $J=1$ lines.
Half the AGN spectra show weaker \Htwo\ absorption
with log~\NHtwo\ $\approx$ 14 -- 17.
A summary of our results appears in Table 2, where
we list \Htwo\ column densities in each rotational state
($J = 0-5$) and the doppler parameter ($b$) from the CoG.

Our analysis routines are adapted from Tumlinson \etal\ (2002), who 
developed our standard software for analyzing \Htwo.  These routines 
allow us to determine equivalent widths for as many \Htwo\ lines 
as possible, and then to apply a curve of growth (CoG) to arrive at  
column densities in individual rotational states, $N(J)$.  From
this information, we can infer various properties of the gas.
We fitted all detected \Htwo\ lines between 1000 and 1126 \AA\ that
are not visibly compromised by continuum problems or line
blending, using either a Gaussian or Voigt profile in
order to measure the equivalent width, \W.  For most
sight lines, the fits to the absorption lines are carefully
tailored on a line-by-line basis to choose the correct
continuum and adjust for any line blending.  When a proper fit
to a line or group of lines has been obtained, the equivalent
width \W, uncertainty in \W, and central wavelength are stored
in an electronic table for use by the CoG fitting routines.
In select cases, we fitted the $J \geq 2$ lines with a more
automated routine, also described in Tumlinson \etal\ (2002), 
which queries the user at each wavelength corresponding to an \Htwo\
line and allows it to be fitted with a single Gaussian component.
This routine is useful for targets with relatively flat continua
and minimal line blending.  The automated routine also stores
\W, uncertainty in \W, and central wavelength in a table.  The
$J=0$ and 1 lines are fitted by hand with a Voigt profile
since they may be strong enough to have non-Gaussian profiles.
For lines between 1036--1062 \AA\ in the Lyman (5-0) and (4-0) bands,
with well-determined centroids, we converted the central wavelength to 
a velocity using the Doppler formula and then averaged the velocities.
We list these \Htwo\ cloud velocities (LSR) in Table 3, together with
velocities of the various 21-cm emission components (Wakker \etal\ 2003).

A single sight line is likely to intercept many clouds containing
\Htwo, that may or may not be offset in velocity.  One of the potential 
sources of systematic uncertainty is our inability to resolve components 
of \Htwo\ absorption by different clouds along the line of sight.  At the 
\FUSE\ resolution of 20 \kms,  we typically cannot separate absorption
components in the ``LSR core'', from $-20$ to $+20$ \kms.  
Components separated by $\geq$ 30 \kms\ can sometimes be resolved, depending 
on the line strengths and widths.   We have 24 targets in common with 
Richter \etal\ (2003), who focused on \Htwo\ in intermediate velocity clouds 
with $|v| = 30-90$ \kms.  They report the existence of an IVC in 7 of our 
sight lines and cite a possible IVC in 6 others.  Because we focus
on low-velocity \Htwo, we generally avoid IVCs in our line fitting.
We have determined the column density for an IVC in three cases: 3C~273,
Mrk~876, and NGC~4151.  For 3C~273, we measured \Htwo\ at  +25 \kms, 
which we do not consider a true IVC.  For Mrk~876, we fitted the IVC lines 
along with the LSR core lines (double component fits). For NGC~4151, the \Htwo\ 
lines arise mainly from IVC gas, but they may have contributions from an LSR 
component.  Out of the 7 targets with confirmed IVCs, it is possible in one 
case that the IVC contributed to our measured LSR \Htwo\ lines.  For the 6 
targets with possible IVCs, it is possible in four cases that an IVC 
contributed to our fitted LSR \Htwo\ lines.  We also note 4 targets where 
Richter \etal\ (2003) claim ``no evidence'' for an IVC, but we believe an 
IVC contributed to our measured \Htwo\ lines.  Our comments on velocity
components and IVCs are summarized
in Table 3.  Multiple unresolved components in the LSR core, as well as 
possible contributions from IVCs, may produce a composite CoG [$b$ and 
$N(J)]$ that reflects the properties of the different velocity components. 

Once the line fitting is complete, the software described in
Tumlinson \etal\ (2002) produces an automated least-squares fit to a CoG 
with a single doppler $b$ parameter for all $J$ levels.  For the two 
sight lines noted above (3C~273 and Mrk 876), we make a two-component
fit for the LSR core and IVC.   Table 2 gives the best-fit values of $N(J)$ 
and single-$b$, with 1 $\sigma$ confidence intervals.  For low-S/N
sight lines or those with low column densities, it is sometimes
difficult to measure enough lines to constrain the CoG.  This
results in large error bars on $N(J)$ and $b$.  In some cases,
model spectra of varying $N(J)$ and $b$ are overlaid on the
spectrum to narrow the range of reasonable values.  In many cases,
the Lyman (1-0) and (0-0) $J = 3$ lines are crucial in constraining
the $b$ value.  Large error bars on $N(J)$ and $b$ often result
when these lines are absent.

For sight lines with no discernable \Htwo\ lines, our methodology
is as follows. We place upper limits on the equivalent widths of
the strong Lyman (7-0) R(0) and R(1) lines and convert these limits
to column density limits assuming a linear CoG.
For the (4~$\sigma$) limiting equivalent width of an unresolved line
at wavelength $\lambda_{0}$, we use the following expression:
\begin{equation}
  W_\lambda^{\rm lim} = \frac {4\lambda_{0}}
                        {(\lambda/\Delta\lambda)(\rm{S/N})} \; .
\end{equation}
Since the resolution of \FUSE\ varies across the band, we
conservatively assume $R = \lambda/\Delta \lambda \approx 15,000$
for all upper limits.  The limiting equivalent widths range from
8~m\AA\ for high-S/N data to 43~m\AA\ for low-S/N data.  This
corresponds to a limiting column density range of
log~\NHtwo\ $> 13.8-14.6$.  The upper limits are included in Table 2.

\subsection{H~I COLUMN DENSITIES }

In order to fully interpret our observations of interstellar \Htwo,
we obtain supplementary data on the atomic hydrogen
(\ion{H}{1}) column densities associated with the observed \Htwo.
Since our sample of 45 sight lines is a subset of that in Wakker \etal\
(2003), we use their compilation of the 21-cm spectrum, as well
as the component structure and N(\ion{H}{1}) determined by fits to the data.
They collected 21-cm spectra from the following sources: Leiden-Dwingeloo Survey
($35'$ beam), Villa Elisa telescope ($34'$ beam), Green Bank 140-ft
telescope ($1'$ beam), and the Effelsberg telescope ($9.'7$ beam).
Wakker \etal\ (2003) provide the 21-cm spectrum, as well
as the component structure and N(\ion{H}{1}) determined by fits to the data.

To derive \NHI, we must determine which of the \ion{H}{1} components are 
physically associated with the observed \Htwo.  This assignment can be 
uncertain and somewhat subjective, so we take care to choose a ``best value"
and a range of \ion{H}{1} column densities, from N$_{\rm min}$ to N$_{\rm max}$.   
Figure 2 illustrates our technique for Mrk~509, with the 21-cm emission
line, \Htwo\ Lyman (4-0) R(0) absorption line, and \ion{Ar}{1} 1048.22 \AA\
absorption line aligned in velocity space. We shift the \FUSE\ data
with respect to the 21-cm data in order to align the \ion{H}{1} emission and
\ion{Ar}{1} absorption as described in \S~2.1.  
We then select as N$_{\rm best}$ the sum of column
densities in the \ion{H}{1} components that are associated with \Htwo\ 
absorption, based on the coincidence of the radial velocities.  Of these 
selected components, at the 20 \kms\ resolution of \FUSE, there remains some 
uncertainty about which are actually associated with \Htwo.
In order to reflect this systematic uncertainty and to be conservative with 
errors, we adopt a range for the \ion{H}{1} column density,
including a floor of systematic uncertainty of $\pm0.03$ in log \NHI.  
For the lower limit, N$_{\rm min}$, we use the smallest individual column density 
that could be responsible for the \Htwo\ absorption within the velocity range 
of the R(0) or R(1) line widths. For the upper limit, N$_{\rm max}$, 
we use the sum of all the H~I column densities that can associated 
kinematically with the \Htwo.  For sight lines
with no discernable \Htwo\ lines, we adopt a lower limit on \NHI\
as the smallest of the components that is not classified
as an IVC or HVC and that satisfies the condition \NHI\ $\ge10^{19}$ \cd.

Further systematic uncertainty arises from the larger beam size of the
21-cm data compared with the pencil-beam sight lines
of the \Htwo\ data.  It has been shown that a smaller beam gives a better
approximation to the \ion{H}{1} column density in the pencil beam (Wakker \&
Schwarz 1991; Wakker \etal\ 2001, 2002).  On the other hand,
Lockman \& Condon (2005), using the $9'.8$ Green Bank telescope, find
that \NHI\ is highly correlated spatially.  
In Table 4, we provide various information on \Htwo\ and H~I, as well
as the 21-cm beam size.


\section{RESULTS}\label{results}

\subsection{General Results}

As noted earlier, \Htwo\ absorption is present in most
of the high-latitude sight lines.
Figure~3 shows a sample \FUSE\ spectrum of PG~1211+143,
an AGN sight line that has been studied extensively with 
both \FUSE\ and {\it Hubble Space Telescope} (HST) by
Penton, Stocke, \& Shull (2004) and Tumlinson \etal\ (2005).
This sight line has a substantial column density,
log~\NHtwo $= 18.38^{+0.15}_{-0.14}$, producing a clear
\Htwo\ band structure throughout the FUV.  The lines are
saturated, but strong damping wings are not yet present.
Figure 4 shows a sequence of four \FUSE\ spectra of AGN,
in order of increasing \Htwo\ column density.  These
range from Ton~S180, where we detect no \Htwo\ to a
($4 \sigma$) limit log \NHtwo\ $<14.37$ ($<13.98$ in $J=0$ and
$<14.14$ in $J=1$), to ESO~141-G55, in which log~\NHtwo\ =
$19.32 \pm 0.07$ with prominent damping wings
in the R(0) and R(1) lines arising from $J = 0$ and $J = 1$.
This montage shows a number of \Htwo\ absorption
lines from $J = 0-4$ in the important Lyman (4-0)
band, along with interstellar lines of Ar~I and Fe~II,
the former of which is used to define the LSR velocity scale. 

Table 2 lists the \Htwo\ column densities in individual rotational 
states, $J = 0-5$ and the doppler ($b$) parameters from the CoGs.
Figure 5 displays the distribution of \Htwo\ column densities,
from the detectable limit, log~\NHtwo\ $\approx 14.2$, up to
the maximum observed value log~\NHtwo\ $\approx 19.8$.  Except for
the dip at log~\NHtwo\ = 17--18, the distribution is
fairly flat in column density.  Small-number statistics
preclude any firm conclusion about whether this dip is real,
but it does suggest two populations of \Htwo\ absorbers.
For each sight line, Table 4 lists the total \Htwo\ and H~I
column densities, molecular fraction ($f_{\rm H2}$),
rotational temperature, $T_{01}$, of $J=0$ and $J=1$
states, and the excitation temperature, $T_{\rm exc}$,
that characterizes the higher-$J$ states.  These temperatures
were derived by least-squares fits of the column densities,
$N(J)$, to the form,
\begin{equation}
  \frac {N(J)} {Z_{\rm H2}} = g_J \exp
           \left( - \frac {E_J}{kT} \right) \; .
\end{equation}
Here, $g_J = g_S (2J+1)$ is the statistical weight of rotational
level $J$, with spin factor $g_S = 1$ or 3 for para- or ortho-\Htwo,
respectively, $E_J$ is the excitation energy of level $J$,
$Z_{\rm H2}$ is the \Htwo\ rotational partition function,
and $T$ denotes $T_{01}$ for rotational states $J$ = 0 and 1
or $T_{\rm exc}$ for excited states $J \geq 2$.

\subsection{Molecular Fraction}

Among the parameters listed in Table~4 is the average molecular
fraction,   
\begin{equation}
   f_{\rm H2} = \frac {\rm{2N(H_2)}} {\rm{N(H~I) + 2N(H_2)} } \; ,
\end{equation}
which expresses the fraction of all hydrogen atoms bound into
\Htwo\ molecules.
We define the total column density of hydrogen as 
N$_{\rm H} \equiv$ \NHI\ + 2\NHtwo, derived from \NHtwo, the \Htwo\ 
column density in all rotational states, and \NHI, the neutral hydrogen 
column density derived from 21-cm emission (\S~2.3).  Error bars on 
log~\NHtwo\ are found by propagating the uncertainties on log~N(0) and 
log~N(1), which dominate the \Htwo\ column density.

In optically thin clouds, the density of molecules can be
approximated by the equilibrium between formation and destruction,
\begin{equation}
 f_{\rm H2} \approx \frac {2 n_H R(T_{\rm gas}, T_{\rm gr}, Z) }
              {\beta \; \langle f_{\rm diss} \rangle }
    \approx (10^{-5}) R_{-17} \; n_{30} \left(
        \frac {\beta_0} {\beta} \right)  \; .
\end{equation}
In this formula, the numerical value for $f_{\rm H2}$ is scaled to
fiducial values of hydrogen density, $n_H$ (30 cm$^{-3}$), \Htwo\
formation rate coefficient, $R$ ($10^{-17}$ cm$^3$~s$^{-1}$), and
mean \Htwo\ pumping rate in the FUV Lyman and Werner bands,
$\beta_0 =  5 \times 10^{-10}$ s$^{-1}$. The \Htwo\ photodissociation
rate is written as $\langle f_{\rm diss} \rangle \beta$, where the
coefficient $\langle f_{\rm diss} \rangle \approx 0.11$ is the
average fraction of FUV excitations of \Htwo\ that result in decays
to the dissociating continuum.  The \Htwo\ formation rate per unit
volume is written as $n_H n_{\rm HI} R$, where $R$
depends on the gas temperature, grain surface temperature, and gas
metallicity ($Z$). The metallicity dependence comes from the assumed
scaling of grain-surface catalysis sites with the grain/gas ratio.
For sight lines in the local Galactic disk, this rate coefficient has
been estimated (Jura 1974) to range from
$R = (1-3) \times 10^{-17}$ cm$^3$~s$^{-1}$ for solar metallicity.
This standard value for $R$ is expected to apply when H~I can stick 
to grain surfaces at suitably
low temperatures of gas ($T_{\rm gas} \leq 300$~K) and grains
($T_{\rm gr} \leq 100$~K) as discussed by Shull \& Beckwith (1982)
and Hollenbach \& McKee (1979).  Although some infalling halo gas may 
have metallicities as low as 10--25\% solar, as in HVC Complex~C 
(Collins \etal\ 2003), we expect that most of the \Htwo-bearing clouds 
in the low halo will have near-solar abundances and grain/gas ratios.

Once sufficient column density of \Htwo\ builds up, with
\NHtwo\ $\geq 10^{15}$ \cd, the cloud becomes optically thick in 
individual Lyman and Werner lines, and the dissociation rate
begins to drop, owing to self-shielding in the lines.  The molecular
fraction rises and makes an abrupt transition to much higher values,
$f_{\rm H2} \geq 10^{-2}$, when the strong R(0) and R(1) lines develop
damping wings at \NHtwo\ $\geq 10^{18}$ \cd.
For our halo survey, Figure 6 shows the molecular fractions, $f_{\rm H2}$,
vs. total H column density. The molecular fraction rises from low values,
$f_{\rm H2} \approx 10^{-5.5 \pm 0.5}$, to values above $10^{-2}$, typical
of sight lines in the disk.  The transition for halo sight lines occurs
over a range log N$_{\rm H} \approx 20.2-20.5$ (we quote a transition at
log~N$_{\rm H} = 20.38 \pm 0.13$), approximately a factor of two below the
similar transition in the Milky Way disk at log~N$_{\rm H} = 20.7$
(Savage \etal\ 1977; Shull \etal\ 2005).
The precise location of the halo transition is not well determined from
just 45 sight lines.  Some broadening of the distribution is expected,
since these sight lines may sample a mixture of gas in the low halo 
(low $f_{\rm H2}$) and denser clouds in the disk (higher $f_{\rm H2}$).

The large observed fluctuations in \Htwo\ column density
require a very patchy ISM in the halo.  For 25 sight lines in the
northern Galactic hemisphere at latitude $|b| \geq 20^{\circ}$,
the mean total \Htwo\ column density is
$\langle {\rm N}_{\rm H2} \sin (b) \rangle = 16.80$,
with a range from $10^{14}$ to $10^{19.8}$~\cd.
In the {\it Copernicus} \Htwo\ survey, Savage \etal\ (1977)
found an average mid-plane density $n_0$(\Htwo) = 0.036 cm$^{-3}$
for 76 stars, which they increased to 0.143 cm$^{-3}$ after
correction for sampling biases due to reddening.
If the \Htwo\ absorption was distributed
smoothly in an exponential layer, $n(z) = n_0 \exp (-z/h)$, with
scale height $h$, the integrated column density toward a quasar at
latitude $b$ would be \NHtwo\ $= n_0 h/ \sin (b)$.
For $h \approx 100$ pc, the column density would exceed
\NHtwo\ = $10^{19}$~\cd, even at high latitude.  Obviously, a
smooth distribution of \Htwo\ is in disagreement with our observations.

The obvious explanation for the wide variations in \NHtwo\ is a clumpy medium.
However, the observed range (and fluctuations) in \NHtwo\ constrain the
geometry and hydrogen density of the \Htwo-bearing ``clouds".  For example,
consider an ensemble of spherical clouds of radius
$r_{\rm cl} = (3~{\rm pc})r_3$ and density $n_H = (30~{\rm cm}^{-3})n_{30}$.
If \Htwo\ is in formation-destruction equilibrium with the mean FUV
radiation field, the molecular density inside the cloud is:
\begin{equation}
   n_{\rm H2} \approx \frac {n_H^2 \; R}{0.11 \beta} \approx
     (1.64 \times 10^{-4}~{\rm cm}^{-3}) n_{30}^2 R_{-17}
      \left( \frac {\beta_0}{\beta} \right) \; ,
\end{equation}
where, as before, we have scaled the \Htwo\ formation rate coefficient
$R$ to $10^{-17}$ cm$^3$~s$^{-1}$.  The typical \Htwo\ column
density through one such cloud would be
\begin{equation}
   N({\rm H}_2) = \frac {4n_{\rm H2} r_{\rm cl} } {3} \approx
       (2 \times 10^{15}~{\rm cm}^{-2}) n_{30}^2 R_{-17}
       \left( \frac {\beta_0}{\beta} \right) ~ r_3 \; .
\end{equation}
In order to explain the total \Htwo\ column density, a sight line to an 
AGN would have to intercept hundreds of such
absorbers.  The fluctuations in the number of clouds along
different sight lines would then be quite small, certainly much less than
the range observed.  On the other hand,  a smaller number of dense,
sheetlike clouds, with a patchy distribution like the infrared cirrus,
would be consistent with the data.  These clouds would each have
more \NHtwo, so the required number of interceptions would be
smaller. Some high-latitude sight lines would have very
large column densities, while others would be quite low, as in the
``\Htwo\ Holes" that we discuss in \S~3.5.  We return to the
cirrus model in \S~3.4.

\subsection{Rotational Temperature}

Figure 7 shows the $J$ = 0--1 rotational excitation
temperature, $T_{01}$, which can measure the kinetic
temperature in dense clouds, where H$^0$--\Htwo\ and
H$^+$--\Htwo\ collisions dominate the mixing of the
$J=0$ and $J=1$ rotational levels.  The solid squares
indicate high-\NHtwo\ sight lines, where models suggest that
the rotational temperature should be close to the gas kinetic
temperature. The quantity $T_{01}$ is obtained from the expression
\begin{equation}
 T_{01} = \frac{\Delta E_{01}/k}{\ln[(g_1/g_0)N(0)/N(1)]} \; ,
\end{equation}
where $g_1/g_0 =9$ is the ratio of the statistical weights of
the $J=1$ and $J=0$ rotational levels, and
$\Delta E_{01}/k =170.5$~K is the excitation temperature of
the $J=1$ level.

To assess the statistical properties of the \Htwo\ rotational
temperature in the halo, we selected 29 sight lines with
sufficiently small error bars on N(0) and N(1) to provide useful
information on $T_{01}$. The halo rotational temperatures range
from $T_{01}$ = 68--252~K, with a median of 139~K, a range that
extends higher than the \FUSE\ Galactic disk values,
all but two of which lie between 55--120~K (Shull \etal\ 2005).
For these 29 sight lines, the mean rotational temperature,
$\langle T_{01} \rangle = 124 \pm 8$~K,
is larger than the values found in the {\it Copernicus} \Htwo\
survey ($77 \pm 17$~K, Savage \etal\ 1977) and in our \FUSE\
Galactic disk survey ($86 \pm 20$~K, Shull \etal\ 2005).
For the 15 sight lines with log~\NHtwo\ $\geq 17$, we find
$\langle T_{01} \rangle = 109 \pm 7$~K.

An elevated kinetic temperature in the low halo could signal  
a shift in thermal balance between UV photoelectric heating by 
dust grains and radiative cooling, primarily by
the [C~II] 158 $\mu$m fine structure line.  Both the heating
and cooling rates should depend on the gas-phase metallicity,
if the grain abundance scales with metallicity.  In the low halo,
we expect $0.1 < (Z/Z_{\odot}) < 1.0$, with most clouds showing 
30\%--100\% solar metallicities.  Even though the abundances for 
these clouds might differ, the equilibrium gas temperature should 
be nearly independent of metallicity (Shull \& Woods 1985),
owing to cancelling dependences of heating and cooling on $Z$.  
The distribution of measured $T_{01}$ is broad, and some of
the sight lines have large error bars.  The median temperature
is 121~K, although a minimum-$\chi^2$ fit gives $83 \pm 6$~K.
Therefore, it may be premature to quote a higher temperature of
these clouds. Additional studies of high-latitude sight lines would be
most helpful.

\subsection{Excitation Temperature}

Figure 8 shows the higher-$J$ excitation temperature, $T_{\rm exc}$,
which reflects non-thermal UV fluorescent pumping of the high-$J$
levels. The mean excitation temperature in the Galactic halo survey
is $\langle T_{\rm exc} \rangle = 505 \pm 28$~K, with a range from
200--852~K.  The median value is 501~K, while a $\chi^2$ fit
to a constant temperature gives $416 \pm 15$~K.   These values are
similar to those seen by the {\it Copernicus} survey (Spitzer \&
Jenkins 1975; Shull \& Beckwith 1982), but slightly higher than that
($326 \pm 125$~K) found in our \FUSE\ disk survey (Shull \etal\ 2005).

Figure 9 shows three ratios of high-$J$ rotational states: N(3)/N(1),
N(4)/N(2), and N(4)/N(0) versus log \NHtwo\ in both low-latitude and
high-latitude \Htwo\ surveys.   The high-latitude survey contains 32 
sight lines with $J=3$ measurements. Of these, only 8 sight lines have
measured values of both N(3) and N(4), and all have substantial
column densities, log~\NHtwo\ $\geq 18$. The $J$-ratios in these absorbers
reflect the relative effects of UV excitation and collisional
de-excitation for para-\Htwo\ ($J = 0$ and 2) and
ortho-\Htwo\ ($J = 1$ and 3), respectively.  For the N(4)/N(2) ratio,
the approximate ranges for the surveys are:
\begin{itemize}

\item {\it High-Latitude: } log N(4)/N(2) = $-2.8$ to $-1.8$ and log \NHtwo\ = 18.5--20.0

\item {\it Low-Latitude: } log N(4)/N(2) = $-4.0$ to $-1.0$ and log \NHtwo\ = 19.5--21.0

\end{itemize}

In general, these two distributions appear to cover similar ranges,
particularly for absorbers with log~\NHtwo\ $\geq 18$.
The greatest distinction between these populations occurs for
the ratio N(3)/N(1) in a subset of 16 high-latitude sight lines with 
low \Htwo\ column density, log~\NHtwo\ = 14--17, and have ratios 
log~N(3)/N(1) = $-1.5$ to $-0.5$, well above the mean value
in the disk ($-3.0$).  These clouds are more optically thin in
the dissociating FUV radiation, and thus have not built up large
column densities in $J = 0$ or 1.   For the 8 sight lines with data for 
both $J=3$ and $J = 4$, the degree of UV excitation appears similar in 
both latitude regimes. This result is consistent with models of
the Galactic radiation field above the disk (Wolfire \etal\ 1995;
Bland-Hawthorn \& Maloney 1999), which suggest that the FUV radiation
field drops by only $\sim50$\% at elevation $z \approx 1$ kpc.    

The high-latitude sight lines typically have a factor of 30 lower \Htwo\ 
column densities than those in the disk. For the disk absorbers, the highly 
excited $J$ populations probably reside in surface layers surrounding a core
with greater column densities in N(0) and N(1).  Our recent models of \Htwo\ 
formation and excitation processes (Browning \etal\ 2003) can be used to 
estimate the effects UV radiation field and gas density from ratios such 
as N(4)/N(2) or N(3)/N(1).  Figure 10 shows models that confirm these 
predictions in our high-latitude data.   

Taken together, these ratio distributions suggest that the degree
of rotational excitation in the high-latitude sight lines is comparable to
that in our disk survey (Shull \etal\ 2005) for high-\NHtwo\ sight
lines.  At first glance, this conclusion may seem inconsistent with the
observed distribution of molecular fractions, $f_{\rm H2}$, with column
density, N$_{\rm H}$.  For the halo sight lines, the transition in $f_{\rm H2}$
occurs at 50\% lower N$_{\rm H}$.  As shown in equation (4), 
the molecular fraction scales as $f_{\rm H2} \propto (n_H  R/ \beta)$.
Thus, a shift in the distribution could arise either from a {\it reduced}
rate of \Htwo\ dissociation (low $\beta$) or an {\it increased}
rate of \Htwo\ formation, $n_H R$, on grain surfaces.  The formation rate
could be increased by a higher grain abundance or surface area (unlikely at 
the metallicities of halo clouds) or by a higher density $n_H$
resulting from cloud compression. Because we cannot supply clear
evidence for an anomalously efficient population of dust grains in
these high-latitude clouds, we favor the high-$n_H$ explanation.
However, additional studies of the dust properties of these absorbers
would be useful.   

Therefore, our best explanation of
the observed $J$-ratios and the shift in the $f_{\rm H2}(N_H)$ transition
is that comparable UV dissociation rates in the halo and disk are
offset by enhanced grain formation of \Htwo\ arising from halo clouds
with higher density than in the disk.  For example, these halo \Htwo\
absorbers may be compressed sheets associated with the IR cirrus, as
suggested for the strong \Htwo\ toward ESO~141-G55 (Shull \etal\ 2000).
We have followed up this idea (Gillmon \& Shull 2005), where we find a
clear association between \Htwo\ and IR cirrus clouds.  Specifically,
we see the same self-shielding transition of $f_{\rm H2}$ versus 100 $\mu$m 
cirrus intensity as seen with N$_{\rm H}$.  This suggests that the cirrus 
clouds with \NHtwo $\geq 10^{18.5}$ \cd\ have a substantial molecular content, 
ranging from 1--30\%.

\subsection{Holes in the \Htwo\ Distribution }

Many of our high-latitude AGN targets lie behind regions of
low \Htwo\ column density.  As noted earlier, most
of our sight lines show Galactic \Htwo\ absorption
with detected column densities ranging from \NHtwo\
$= 10^{14.17-19.67}$~\cd.   In the northern Galactic hemisphere,
8 sight lines have weak \Htwo\ absorption (\NHtwo\ $\leq 10^{15}$ \cd)
located between $l=60^{\circ} - 180^{\circ}$ at
$b > 54^{\circ}$.  We have identified these high-latitude regions
as ``H$_2$ holes", a patchy network of interstellar gas with both
low 21-cm emission and low \Htwo\ absorption (Figure 11).

The 8 sight lines through the \Htwo\ holes provide special locations to
probe diffuse halo gas, including large-scale halo structures, IVCs, and HVCs. 
The structure of these holes may have been shaped by
infalling H~I clouds, as well as by outflowing gas from the ``Northern
Chimney", an interstellar cavity inferred from low Na~I absorption at
$z \leq 300$~pc (Lallement \etal\ 2003).
Large portions of the holes have \NHtwo\ $\leq 10^{15}$~\cd.
These regions of low \NHtwo\ are analogous to the ``Lockman Hole"
of low \NHI\ (Lockman \etal\ 1986), but they have much greater extent.
Figure 11 compares our \Htwo\ survey with the distribution of \NHI\ in 
the northern Galactic hemisphere at $b \geq 45^{\circ}$.


\section{DISCUSSION AND SUMMARY}

Our \FUSE\ high-latitude survey was able to probe molecular fractions below
and above the transition value at log~N$_{\rm H} \approx 20.38 \pm 0.13$.
We compared this transition to that seen in
the Galactic disk surveys (Savage \etal\ 1977; Shull \etal\ 2005).
Our sample, and possibly an enlarged future survey, can be used
to examine fundamental properties of \Htwo, tied to O~VI, H~I, and
IR cirrus, in an optically thin regime (log~\NHtwo\ $<17$)
inaccessible to \FUSE\ in the Galactic disk.  These low-column
clouds provide important tests of our theoretical models of
\Htwo\ formation, destruction, and excitation in optically-thin
environments, where FUV radiation dominates the $J$-level populations.

These high-latitude sight lines are also relevant for far-UV studies
of interstellar gas, intergalactic gas, and HVCs, since they provide
``\Htwo-clean" sight lines with minimal contamination.
In the future, with \FUSE\ or a next-generation satellite
with FUV capability, these high-latitude sight lines through the
\Htwo\ holes should be prime targets for studies of HVCs, gas at the
disk-halo interface, and the IGM.
They can be used to address several important astronomical issues:
(1) A possible  correlation of \Htwo\ holes with enhanced O~VI absorption;
(2) The correlation of \Htwo\ with H~I and infrared cirrus structures;
(3) High-latitude probes of HVCs, IVCs, and Galactic fountain gas.

As discussed in \S~3.4, the distribution of molecular fractions,
$f_{\rm H2}$, and the level of rotational excitation to levels
$J \geq 2$, can be used to infer the FUV radiation field
(\Htwo\ pumping rate $\beta$) and gas density in the absorbers.
Because $f_{\rm H2} \propto (n_H R/\beta)$ in formation-destruction
equilibrium, the high-latitude sight lines with high $f_{\rm H2}$ require 
either reduced $\beta$ or an enhanced formation rate, $n_H R$.
The magnitude of the \Htwo\ pumping rate ($\beta$) was assessed
independently by examining populations of the high-$J$ rotational states.
Figures 9 and 10 show that the ratios, N(3)/N(1), N(4)/N(0), and N(4)/N(2),
for high-latitude sight lines with \NHtwo\ $\geq 10^{18}$ \cd\ are not 
greatly different from those in the Galactic disk.  Thus, it is difficult 
to argue that the FUV radiation field at $z \leq 200$ pc is significantly
lower than in the disk.  Within a factor of two, the absorbers probed in 
this survey seem
to have $\beta \approx \beta_0$, the mean value in the Galactic disk.
More distant clouds ($z \geq 1$ kpc) may experience a lower FUV radiation 
field, as they are farther from OB stars in the Galactic disk.  

Higher \Htwo\ formation rates could arise from more efficient grain
catalysis (higher $R$ arising from increased grain/gas ratio) or from 
denser gas (higher $n_H$).  At $z < 200$ pc, the dust/gas ratio probably 
tracks the metallicity and is unlikely to differ much from that in the 
disk.  It is more plausible that clouds at the disk-halo interface have 
been compressed dynamically and have densities larger, on average, than 
quiescent diffuse clouds in the disk.  We therefore favor a model in which 
the observed shift in the distribution of $f_{\rm H2}$ with N$_{\rm H}$ arises 
from more efficient \Htwo\ formation rate in high-density cirrus clouds.

In a separate paper (Gillmon \& Shull 2005) we identified a
correlation between high-latitude \Htwo\ and infrared cirrus.
This correlation means that 100~$\mu$m cirrus maps (Schlegel, Finkbeiner,
\& Davis 1998) can be used to identify the best regions to
explore low-\NHtwo\ sight lines in future FUV studies.  This
technique also allows us to characterize the physical properties
and total molecular mass of the halo clouds.  Models of the
rotational excitation can be used to estimate the FUV radiation
field and gas densities, while the IR cirrus can be used to
constrain grain temperatures (60/100~$\mu$m ratios), define
the cloud spatial extent, and characterize the
differences from clouds in the Galactic disk.  By measuring
the gas metallicity and dust depletion from UV resonance lines,
one may even be able to draw connections between the \Htwo, gas
metallicity, and dust heating rate.

\acknowledgments{ }

We thank James Green, Bart Wakker, Blair Savage, and Ken Sembach for 
useful discussions, and Matthew Browning for his assistance with
the modeling of \Htwo\ rotational populations.  
We are grateful to the referee, Phillipp Richter,
for valuable comments that helped clarify the importance on \Htwo\
in IVCs and the disk-halo interface.  This work was based in part 
on data obtained for
the Guaranteed Time Team team by the NASA-CNES-CSA \FUSE\ mission
operated by the Johns Hopkins University.  Financial support to
U.S. participants has been provided by NASA contract NAS5-32985.
The Colorado group also received \FUSE\ support from
NASA grant NAG5-10948 for studies of interstellar \Htwo.

\clearpage


\clearpage


\begin{deluxetable}{lcccccccc}
\tabletypesize{\scriptsize}
\tablecolumns{9}
\tablenum{1}
\tablewidth{0pt}
\tablecaption{GALACTIC HALO TARGET LIST}

\tablehead{
\colhead{Name}
&\colhead{$l$}
&\colhead{$b$}
&\colhead{Type}
&\colhead{B}
&\colhead{$E(B-V)$\tablenotemark{a}}
&\colhead{Observation ID}
&\colhead{$t_{\rm exp}$\tablenotemark{b}}
&\colhead{S/N\tablenotemark{c}}\\
\colhead{}
&\colhead{(deg)}
&\colhead{(deg)}
&\colhead{}
&\colhead{}
&\colhead{}
&\colhead{}
&\colhead{(ks)}
&\colhead{($\rm{pixel^{-1}}$)}
}

\startdata
{3C 249.1} &130.39 &38.55 &QSO &15.70 &0.034 &P1071601 &11.4 &2\\
{} & & & & & &P1071602 &9.5 &\\
{} & & & & & &P1071603 &80.9 &\\
{} & & & & & &D1170101 &(20.8) &\\
{} & & & & & &S6010901 &30.7 &\\
{3C 273} &289.95 &64.36 &QSO &13.05 &0.021 &P1013501 &43.2 &10\\
{ESO 141$-$G55} &338.18 &-26.71 &Sey1 &13.83 &0.111 &I9040104 &(40.4) &3\\
{H 1821+643} &94.00 &27.42 &QSO &14.23 &0.043 &P1016402 &62.6 &4\\
{} & & & & & &P1016405 &23.1 &\\
{HE 0226$-$4110} &253.94 &-65.77 &QSO &14.30 &0.016 &P2071301 &11.0 &4\\
{} & & & & & &P1019101 &64.3 &\\
{} & & & & & &P1019104 &18.1 &\\
{} & & & & & &D0270101 &(23.9) &\\
{} & & & & & &D0270102 &(39.9) &\\
{} & & & & & &D0270103 &(17.2) &\\
{HE 1143$-$1810} &281.85 &41.71 &Sey1 &14.63 &0.039 &P1071901 &7.3 &2\\
{HS 0624+6907} &145.71 &23.35 &QSO &14.64 &0.098 &P1071001 &13.9 &2\\
{} & & & & & &P1071002 &13.6 &\\
{} & & & & & &S6011201 &43.8 &\\
{} & & & & & &S6011202 &40.6 &\\
{MRC 2251$-$178} &46.20 &-61.33 &QSO &14.99 &0.039 &P1111010 &54.1 &3\\
{Mrk9} &158.36 &28.75 &Sey1.5 &14.77 &0.059 &P1071101 &11.7 &2\\
{}& & & & & &P1071102 &14.2 &\\
{}& & & & & &P1071103 &2.6 &\\
{}& & & & & &S6011601 &23.9 &\\
{Mrk106} &161.14 &42.88 &Sey1 &16.54 &0.028 &C1490501 &(117.1) &3\\
{Mrk 116} &160.53 &44.84 &BCG &15.50 &0.032 &P1080901 &58.4 &7\\
{} & & & & & &P1980102 &31.6 &\\
{Mrk205} &125.45 &41.67 &Sey1 &15.64 &0.042 &D0540101 &(20.7) &3\\
{} & & & & & &D0540102 &(112.4) &\\
{} & & & & & &D0540103 &(20.9) &\\
{Mrk 209} &134.15 &68.08 &BCD &15.22 &0.015 &Q2240101 &19.2 &3\\
{} & & & & & &P1072201 &6.3 &\\
{Mrk 290} &91.49 &47.95 &Sey1 &15.56 &0.015 &P1072901 &12.7 &2\\
{} & & & & & &D0760101 &(9.3) &\\
{Mrk 335} &108.76 &-41.42 &Sey1.2 &14.19 &0.035 &P1010203 &31.7 &11\\
{} & & & & & &P1010204 &53.2 &\\
{Mrk 421} &179.83 &65.03 &BLLac &13.50 &0.015 &P1012901 &21.8 &8\\
{} & & & & & &Z0100101 &(26.5) &\\
{} & & & & & &Z0100102 &(23.4) &\\
{} & & & & & &Z0100103 &(12.1) &\\
{Mrk 478} &59.24 &65.03 &Sey1 &14.91 &0.014 &P1110909 &14.2 &2\\
{Mrk 501} &63.60 &38.86 &BLLac &14.14 &0.019 &C0810101 &(18.4) &2\\
{} & & & & & &P1073301 &11.4 &\\
{Mrk 509} &35.97 &-29.86 &Sey1.2 &13.35 &0.057 &P1080601 &62.3 &6\\
{Mrk 817} &100.30 &53.48 &Sey1.5 &14.19 &0.007 &P1080401 &12.1 &7\\
{} & & & & & &P1080402 &12.8 &\\
{} & & & & & &P1080403 &71.0 &\\
{} & & & & & &P1080404 &86.8 &\\
{Mrk 876} &98.27 &40.38 &Sey1 &16.03 &0.027 &P1073101 &49.4 &6\\
{Mrk 1095} &201.69 &-21.13 &Sey1 &14.30 &0.128 &P1011201 &8.5 &3\\
{} & & & & & &P1011202 &20.8 &\\
{} & & & & & &P1011203 &26.8 &\\
{Mrk 1383} &349.22 &55.12 &Sey1 &15.21 &0.032 &P1014801 &24.8 &6\\
{} & & & & & &P2670101 &38.7 &\\
{Mrk 1513} &63.67 &-29.07 &Sey1 &14.92 &0.044 &P1018301 &22.6 &2\\
{MS 0700.7+6338} &152.47 &25.63 &Sey1 &14.51 &0.051 &P2072701 &7.4 &2\\
{} & & & & & &S6011501 &16.9 &\\
{NGC 985} &180.84 &-59.49 &Sey1 &14.64 &0.033 &P1010903 &50.6 &4\\
{NGC 1068} &172.10 &-51.93 &Sey1 &11.70 &0.034 &P1110202 &22.6 &5\\
{NGC 1705} &261.08 &-38.74 &GAL &11.50 &0.008 &A0460102 &(8.6) &9\\
{} & & & & & &A0460103 &(15.4) &\\
{NGC 4151} &155.08 &75.06 &Sey1.5 &12.56 &0.028 &C0920101 &(54.3) &9\\
{NGC 4670} &212.69 &88.63 &BCD &14.10 &0.015 &B0220301 &(8.7) &5\\
{} & & & & & &B0220302 &(16.0) &\\
{} & & & & & &B0220303 &(10.7) &\\
{NGC 7469} &83.10 &-45.47 &Sey1.2 &13.42 &0.069 &P1018703 &37.3 &6\\
{PG 0804+761} &138.28 &31.03 &QSO &15.03 &0.035 &P1011901 &39.3 &8\\
{}  & & & & & &P1011903 &20.8 &\\
{}  & & & & & &S6011001 &58.5 &\\
{}  & & & & & &S6011002 &36.9 &\\
{PG 0844+349} &188.56 &37.97 &Sey1 &14.83 &0.037 &P1012002 &31.7 &4\\
{PG 0953+414} &179.79 &51.71 &QSO &15.29 &0.013 &P1012201 &36.2 &5\\
{} & & & & & &P1012202 &38.1 &\\
{PG 1116+215} &223.36 &68.21 &QSO &14.85 &0.023 &P1013101 &11.0 &6\\
{} & & & & & &P1013102 &10.8 &\\
{} & & & & & &P1013103 &8.0 &\\
{} & & & & & &P1013104 &10.9 &\\
{} & & & & & &P1013105 &33.5 &\\
{PG 1211+143} &267.55 &74.32 &Sey1 &14.46 &0.035 &P1072001 &52.3 &5\\
{PG 1259+593} &120.56 &58.05 &QSO &15.84 &0.008 &P1080101 &52.3 &7\\
{} & & & & & &P1080102 &48.2 &\\
{} & & & & & &P1080103 &51.8 &\\
{} & & & & & &P1080104 &106.3 &\\
{} & & & & & &P1080105 &104.4 &\\
{} & & & & & &P1080106 &62.9 &\\
{} & & & & & &P1080107 &95.1 &\\
{} & & & & & &P1080108 &32.6 &\\
{} & & & & & &P1080109 &29.8 &\\
{} & & & & & & & &\\
{PG 1302$-$102} &308.59 &52.16 &QSO &15.18 &0.043 &P1080201 &31.9 &4\\
{} & & & & & &P1080202 &32.3 &\\
{} & & & & & &P1080203 &81.9 &\\
{PKS 0405$-$12} &204.93 &-41.76 &QSO &15.09 &0.058 &B0870101 &(71.1) &4\\
{PKS 0558$-$504} &257.96 &-28.57 &QSO &15.18 &0.044 &P1011504 &45.1 &5\\
{} & & & & & &C1490601 &(48.0) &\\
{PKS 2005$-$489} &350.37 &-32.60 &BLLac &13.40 &0.056 &P1073801 &12.3 &5\\
{} & & & & & &C1490301 &(22.5) &\\
{} & & & & & &C1490302 &(13.3) &\\
{PKS 2155$-$304} &17.73 &-52.25 &BLLac &13.36 &0.022 &P1080701 &19.2 &10\\
{} & & & & & &P1080703 &65.4 &\\
{} & & & & & &P1080705 &38.6 &\\
{Ton S180} &139.00 &-85.07 &Sey1.2 &14.60 &0.014 &P1010502 &16.9 &4\\
{Ton S210} &224.97 &-83.16 &QSO &15.38 &0.017 &P1070302 &41.0 &5\\
{VII Zw 118} &151.36 &25.99 &Sey1 &15.29 &0.038 &P1011604 &67.2 &4\\
{} & & & & & &P1011605 &25.9 &\\
{} & & & & & &S6011301 &49.9 &\\
\enddata

\tablenotetext{a}{$E(B-V)$ values are from Schlegel \etal\ (1998) by way of NED
(http://nedwww.ipac.caltech.edu).  Since these values are inferred from IRAS
dust maps for a nonredshifted elliptical galaxy, they may not be appropriate
for other source types.  Thus, these values should be treated as an indication of
relative dust column along the sight line as opposed to the actual reddening for these
AGN sources.}
\tablenotetext{b}{Exposure times are estimates of the actual post-calibration
exposure times.  The exposure times were partially corrected by subtracting the
duration of the subexposures during which the detector was off.  Because the
calibration may omit additional subexposures with low flux or burst activity,
the listed exposure times are upper limits on the actual exposure times.  For
those targets without available subexposure information, the full uncorrected
exposure time is given in parentheses.}
\tablenotetext{c}{For targets with more than one observation, the S/N for the
coadded dataset (including all listed observations) is given in the row with the
target name.}
\end{deluxetable}


\begin{deluxetable}{lccccccc}
\tabletypesize{\scriptsize}
\tablecolumns{8}
\tablenum{2}
\tablewidth{0pt}
\tablecaption{ROTATIONAL LEVEL POPULATIONS}

\tablehead{
\colhead{Name}
&\colhead{log N(0)}
&\colhead{log N(1)}
&\colhead{log N(2)}
&\colhead{log N(3)}
&\colhead{log N(4)}
&\colhead{log N(5)}
&\colhead{$b_{\rm{dopp}}$\tablenotemark{a}}\\
\colhead{}
&\colhead{(\cd)}
&\colhead{(\cd)}
&\colhead{(\cd)}
&\colhead{(\cd)}
&\colhead{(\cd)}
&\colhead{(\cd)}
&\colhead{(\kms)}
}

\startdata
{3C 249.1} &18.40 $\pm^{0.22}_{0.28}$ &18.84 $\pm^{0.14}_{0.17}$ &17.03 $\pm^{0.80}_{0.98}$ &16.41 $\pm^{1.19}_{0.55}$ &-- &-- &6.1 $\pm^{2.0}_{2.3}$\\
{3C 273 ($v=-5$)} &$\le$13.51 &14.30 $\pm^{0.05}_{0.05}$ &-- &-- &-- &-- &linear\\
{\hspace{9.3mm} ($v=26$)} &14.89 $\pm^{0.12}_{0.09}$ &15.50 $\pm^{0.21}_{0.14}$ &14.87 $\pm^{0.10}_{0.07}$ &14.71 $\pm^{0.07}_{0.05}$ &-- &-- &4.9 $\pm^{0.9}_{0.6}$\\
{ESO 141$-$G55} &18.90 $\pm^{0.11}_{0.11}$ &19.10 $\pm^{0.08}_{0.08}$ &17.34 $\pm^{0.44}_{0.79}$ &16.10 $\pm^{1.26}_{0.52}$ &14.64 $\pm^{1.56}_{0.25}$  &-- &3.8 $\pm^{1.4}_{2.4}$\\
{H 1821+643} &16.97 $\pm^{0.26}_{0.40}$ &17.71 $\pm^{0.09}_{0.09}$ &17.18 $\pm^{0.16}_{0.23}$ &16.78 $\pm^{0.31}_{0.61}$ &-- &-- &1.7 $\pm^{0.8}_{0.7}$\\
{HE 0226$-$4110}\tablenotemark{b}
 &$\le$13.90 &$\le$14.06 &-- &-- &-- &-- &--\\
{HE 1143$-$1810} &15.91 $\pm^{1.53}_{0.80}$ &16.37 $\pm^{1.30}_{0.80}$ &15.35 $\pm^{1.34}_{0.31}$ &14.93 $\pm^{0.97}_{0.25}$ &-- &-- &7.0 $\pm^{3.6}_{3.2}$\\
{HS 0624+6907} &19.39 $\pm^{0.13}_{0.14}$ &19.60 $\pm^{0.12}_{0.12}$ &17.95 $\pm^{0.31}_{1.56}$ &17.13 $\pm^{0.57}_{1.13}$ &15.49 $\pm^{0.33}_{0.37}$ &14.63 $\pm^{0.05}_{0.09}$ &6.5 $\pm^{4.2}_{1.0}$\\
{MRC 2251$-$178} &$\le$14.12 &14.54 $\pm^{0.23}_{0.17}$ &-- &-- &-- &-- &linear\\
{Mrk 9} &18.87 $\pm^{0.15}_{0.14}$ &19.18 $\pm^{0.11}_{0.11}$ &17.13 $\pm^{0.35}_{0.46}$ &15.40 $\pm^{0.20}_{0.20}$ &-- &-- &5.4 $\pm^{0.6}_{0.4}$\\
{Mrk 106} &15.95 $\pm^{0.28}_{0.20}$ &15.82 $\pm^{0.22}_{0.18}$ &14.98 $\pm^{0.14}_{0.11}$ &14.73 $\pm^{0.16}_{0.15}$ &-- &-- &14.9 $\pm^{3.5}_{2.6}$\\
{Mrk 116} &18.82 $\pm^{0.13}_{0.13}$ &18.73 $\pm^{0.09}_{0.11}$ &17.12 $\pm^{0.47}_{0.85}$ &16.89 $\pm^{0.58}_{0.71}$ &14.76 $\pm^{0.40}_{0.24}$ &-- &4.5 $\pm^{1.6}_{1.5}$\\
{Mrk 205} &15.63 $\pm^{0.00}_{0.29}$ &16.39 $\pm^{0.00}_{0.45}$ &15.44 $\pm^{0.00}_{0.28}$ &15.40 $\pm^{0.00}_{0.38}$ &-- &-- &7.0 $\pm^{1.8}_{0.0}$\\
{Mrk 209} &$\le$14.10 &$\le$14.25 &-- &-- &-- &-- &--\\
{Mrk 290} &15.80 $\pm^{1.96}_{0.61}$ &15.78 $\pm^{2.03}_{0.39}$ &15.26 $\pm^{2.21}_{0.27}$ &15.03 $\pm^{2.28}_{0.26}$ &-- &-- &8.7 $\pm^{3.7}_{6.0}$\\
{Mrk 335} &18.44 $\pm^{0.11}_{0.12}$ &18.59 $\pm^{0.06}_{0.07}$ &16.78 $\pm^{0.50}_{0.44}$ &16.07 $\pm^{0.51}_{0.24}$ &14.23 $\pm^{0.15}_{0.12}$ &-- &5.8 $\pm^{0.9}_{1.0}$\\
{Mrk 421} &13.89 $\pm^{0.25}_{0.25}$ &14.40 $\pm^{0.13}_{0.09}$ &14.01 $\pm^{0.25}_{0.25}$ &-- &-- &-- &linear\\
{Mrk 478} &$\le$14.18 &$\le$14.33 &-- &-- &-- &-- &--\\
{Mrk 501} &$\le$14.20 &14.78 $\pm^{0.18}_{0.10}$ &-- &-- &-- &-- &linear\\
{Mrk 509} &17.34 $\pm^{0.34}_{0.83}$ &17.69 $\pm^{0.24}_{0.70}$ &16.37 $\pm^{0.62}_{0.61}$ &15.63 $\pm^{0.38}_{0.26}$ &-- &-- &5.8 $\pm^{1.9}_{1.0}$\\
{Mrk 817} &$\le$13.64 &$\le$13.80 &-- &-- &-- &-- &--\\
{Mrk 876 ($v=-33$)} &15.18 $\pm^{2.18}_{0.48}$ &15.53 $\pm^{2.22}_{0.35}$ &14.61 $\pm^{2.22}_{0.26}$ &14.50 $\pm^{2.15}_{0.25}$ &-- &-- &5.7 $\pm^{10.4}_{4.7}$\\
{\hspace{11mm} ($v=-3$)} &15.81 $\pm^{2.08}_{0.50}$ &16.41 $\pm^{1.96}_{0.46}$ &15.59 $\pm^{2.11}_{0.38}$ &15.22 $\pm^{2.23}_{0.30}$ &-- &-- &7.3 $\pm^{3.3}_{5.4}$\\
{Mrk 1095} &18.24 $\pm^{0.25}_{0.38}$ &18.58 $\pm^{0.20}_{0.32}$ &17.10 $\pm^{0.77}_{0.72}$ &16.49 $\pm^{0.92}_{0.50}$ &14.90 $\pm^{0.22}_{0.18}$ &-- &9.2 $\pm^{2.1}_{2.1}$\\
{Mrk 1383} &$\le$13.77 &14.35 $\pm^{0.13}_{0.10}$ &-- &-- &-- &-- &linear\\
{Mrk 1513} &16.12 $\pm^{1.22}_{0.33}$ &16.02 $\pm^{1.06}_{0.28}$ &15.23 $\pm^{0.54}_{0.23}$ &14.87 $\pm^{0.27}_{0.19}$ &-- &-- &11.7 $\pm^{3.6}_{4.1}$\\
{MS 0700.7+6338} &18.41 $\pm^{0.29}_{0.80}$ &18.46 $\pm^{0.25}_{0.65}$ &17.27 $\pm^{0.86}_{1.57}$ &15.37 $\pm^{1.75}_{0.41}$ &-- &-- &7.6 $\pm^{4.5}_{3.6}$\\
{NGC 985} &15.57 $\pm^{2.04}_{0.46}$ &15.80 $\pm^{1.98}_{0.38}$ &14.93 $\pm^{1.86}_{0.20}$ &14.67 $\pm^{1.06}_{0.20}$ &-- &-- &8.1 $\pm^{3.0}_{5.0}$\\
{NGC 1068} &17.84 $\pm^{0.08}_{0.09}$ &17.82 $\pm^{0.08}_{0.11}$ &15.84 $\pm^{0.66}_{0.40}$ &15.39 $\pm^{0.60}_{0.26}$ &-- &-- &3.4 $\pm^{0.8}_{0.8}$\\
{NGC 1705} &$\le$13.78 &$\le$13.94 &-- &-- &-- &-- &--\\
{NGC 4151} &15.95 $\pm^{0.72}_{0.17}$ &16.55 $\pm^{0.93}_{0.34}$ &15.57 $\pm^{0.57}_{0.12}$ &15.19 $\pm^{0.26}_{0.11}$ &-- &-- &7.2 $\pm^{1.1}_{1.9}$\\
{NGC 4670} &14.31 $\pm^{0.25}_{0.24}$ &14.51 $\pm^{0.18}_{0.15}$ &-- &-- &-- &-- &linear\\
{NGC 7469} &19.41 $\pm^{0.10}_{0.09}$ &19.32 $\pm^{0.08}_{0.08}$ &17.77 $\pm^{0.24}_{0.35}$ &16.39 $\pm^{0.75}_{0.35}$ &14.90 $\pm^{0.27}_{0.14}$ &-- &5.5 $\pm^{1.1}_{1.3}$\\
{PG 0804+761} &18.08 $\pm^{0.16}_{0.20}$ &18.52 $\pm^{0.08}_{0.11}$ &16.63 $\pm^{0.54}_{0.37}$ &15.86 $\pm^{0.32}_{0.16}$ &-- &-- &8.7 $\pm^{1.2}_{1.2}$\\
{PG 0844+349} &17.64 $\pm^{0.21}_{0.29}$ &18.09 $\pm^{0.11}_{0.16}$ &16.04 $\pm^{0.78}_{0.40}$ &15.04 $\pm^{0.26}_{0.21}$ &-- &-- &4.7 $\pm^{1.2}_{1.0}$\\
{PG 0953+414} &14.10 $\pm^{0.25}_{0.25}$ &14.76 $\pm^{0.19}_{0.11}$ &14.09 $\pm^{0.25}_{0.25}$ &14.39 $\pm^{0.23}_{0.20}$ &-- &-- &linear\\
{PG 1116+215} &$\le$13.78 &$\le$13.93 &-- &-- &-- &-- &--\\
{PG 1211+143} &17.80 $\pm^{0.13}_{0.16}$ &18.23 $\pm^{0.08}_{0.08}$ &16.72 $\pm^{0.57}_{0.59}$ &15.77 $\pm^{0.81}_{0.25}$ &-- &-- &4.3 $\pm^{1.0}_{1.0}$\\
{PG 1259+593} &13.77 $\pm^{0.25}_{0.25}$ &14.34 $\pm^{0.15}_{0.11}$ &14.06 $\pm^{0.25}_{0.25}$ &14.23 $\pm^{0.25}_{0.25}$ &-- &-- &linear\\
{PG 1302$-$102} &14.79 $\pm^{0.86}_{0.21}$ &15.37 $\pm^{1.64}_{0.23}$ &14.84 $\pm^{0.50}_{0.19}$ &14.71 $\pm^{0.43}_{0.18}$ &-- &-- &11.0 $\pm^{3.0}_{6.0}$\\
{PKS 0405$-$12} &14.91 $\pm^{0.17}_{0.15}$ &15.56 $\pm^{0.27}_{0.14}$ &14.93 $\pm^{0.10}_{0.08}$ &14.96 $\pm^{0.13}_{0.10}$ &-- &-- &11.8 $\pm^{3.1}_{2.6}$\\
{PKS 0558$-$504} &14.51 $\pm^{0.16}_{0.14}$ &15.22 $\pm^{0.19}_{0.14}$ &14.65 $\pm^{0.12}_{0.11}$ &14.52 $\pm^{0.16}_{0.16}$ &-- &-- &10.5 $\pm^{5.4}_{2.7}$\\
{PKS 2005$-$489} &14.32 $\pm^{0.25}_{0.25}$ &14.74 $\pm^{0.18}_{0.09}$ &14.36 $\pm^{0.25}_{0.25}$ &14.29 $\pm^{0.25}_{0.25}$ &-- &-- &linear\\
{PKS 2155$-$304} &13.59 $\pm^{0.25}_{0.25}$ &14.04 $\pm^{0.15}_{0.13}$ &-- &-- &-- &-- &linear\\
{Ton S180} &$\le$13.98 &$\le$14.14 &-- &-- &-- &-- &--\\
{Ton S210} &16.00 $\pm^{1.21}_{1.18}$ &16.44 $\pm^{1.01}_{1.34}$ &-- &-- &-- &-- &4.4 $\pm^{5.6}_{3.4}$\\
{VII Zw 118} &18.38 $\pm^{0.11}_{0.13}$ &18.65 $\pm^{0.07}_{0.08}$ &16.22 $\pm^{0.59}_{0.34}$ &16.01 $\pm^{0.31}_{0.18}$ &14.40 $\pm^{0.16}_{0.15}$ &-- &6.5 $\pm^{1.2}_{1.1}$\\
\enddata
\tablenotetext{a}{Entry of ``linear" means that lines fall on the linear part of the curve of growth, and thus there is no constraint on doppler $b$.}
\tablenotetext{b}{Savage \etal\ (2005) have detected \Htwo\ in this sightline,
with log~N(0) = $13.85\pm0.3$, log~N(1) = $14.06\pm0.3$, log~N(2) = $13.44^{+0.5}_{-1.2}$,
and log~N(3) = $13.51^{+0.5}_{-1.2}$.}
\end{deluxetable}

\clearpage


\begin{deluxetable}{lcccccc}
\tabletypesize{\scriptsize}
\tablecolumns{7}
\tablenum{3}
\tablewidth{0pt}
\tablecaption{\Htwo\ and H~I Velocity Components (LSR)}

\tablehead{
\colhead{Name}
&\colhead{$v$(H~I)\tablenotemark{a}}
&\colhead{$v_{\rm min}$(H~I)\tablenotemark{b}}
&\colhead{$v_{\rm max}$(H~I)\tablenotemark{c}}
&\colhead{$v$(\Htwo)\tablenotemark{d}}
&\colhead{IVC, \Htwo\ status\tablenotemark{e}}
&\colhead{IVC\tablenotemark{f}}\\
\colhead{}
&\colhead{(\kms)}
&\colhead{(\kms)}
&\colhead{(\kms)}
&\colhead{(\kms)}
&\colhead{(\kms)}
&\colhead{Notes}
}

\startdata
{3C 249.1} &$-50,-20,(1,9)$ &9 &$-50,-20,1,9$ &12 &$-50$,Possibly &2\\
{3C 273} &$(-13,-6),0,25$ &$-6$ &$-13,-6,0$ &$-5$ &25,Yes &1\\
{} &$-13,-6,0,(25)$ &25 &25 &26 &25,Yes &3\\
{ESO 141$-$G55} &$-45,-23,(-1,2),30$ &$-1$ &$-45,-23,-1,2,30$ &$-3$ &$-45$,None &2\\
{H 1821+643} &$-87,-23,(-10,-3)$ &$-3$ &$-23,-10,-3$ &$-3$ &-- &--\\
{HE 0226$-$4110} &-- &-- &-- &-- &-- &--\\
{HE 1143$-$1810} &$(-6,-5,4),13,56$ &4 &$-6,-5,4,13$ &3 &-- &--\\
{HS 0624+6907} &$-27,(-10,-3,0,10)$ &10 &$-27,-10,-3,0,10$ &6 &-- &--\\
{MRC 2251$-$178} &$-8,(0),59$ &$-8$ &$-8,0$ &$-8$ &-- &--\\
{Mrk 9} &$-40,(-8,-3,-1,4)$ &$-8$ &$-40,-8,-3,-1,4$ &$-2$ &$-40$,Possibly &2\\
{Mrk 106} &$-40,(-6,-2,-2)$ &$-6$ &$-6,-2,-2$ &$-6$ &$-40$,Possibly &1\\
{Mrk 116} &$-39,-16,(1,7)$ &7 &$-39,-16,1,7$ &4 &$-39$,Possibly &2\\
{Mrk 205} &$-48,(-4,2)$ &2 &$-4,2$ &3 &$-48$,Possibly &1\\
{Mrk 209} &-- &-- &-- &-- &$-55$,None &--\\
{} & & & & &$-100$,None &--\\
{Mrk 290} &$(-6,-5_n,-5_b)$\tablenotemark{g} &$-5_n$ &$-6,-5_n,-5_b$ &$-8$ &-- &--\\
{Mrk 335} &$-27,(-9,-3,-2)$ &$-2$ &$-27,-9,-3,-2$ &$-1$ &$-27$,Possibly &2\\
{Mrk 421} &$-61,(-23,-11,-8)$ &$-23$ &$-23,-11,-8$ &$-15$ &$-61$,None &1\\
{Mrk 478} &-- &-- &-- &-- &-- &--\\
{Mrk 501} &$-83,-38,(-3,-1,2)$ &2 &$-3,-1,2$ &$-1$ &$-38$,None &1\\
{Mrk 509} &$(-7,6),60$ &$-7$ &$-7,6$ &2 &60,Yes &1\\
{Mrk 817} &-- &-- &-- &-- &$-40$,None &--\\
{Mrk 876} &$-52,(-30),-7$ &$-30$ &$-30$ &$-33$ &$-30$,Yes &3\\
{} &$-52,-30,(-7)$ &$-7$ &$-7$ &$-3$ &$-30$,Yes &1\\
{Mrk 1095} &$-13,(4,9)$ &9 &$-13,4,9$ &16 &-- &--\\
{Mrk 1383} &$(-7,-2)$ &$-2$ &$-7,-2$ &0 &-- &--\\
{Mrk 1513} &$-29,(1,5,10)$ &10 &$-29,1,5,10$ &6 &$-29$,None &2\\
{MS 0700.7+6338} &$-11,(-6,-1,6)$ &$-6$ &$-11,-6,-1,6$ &6 &-- &--\\
{NGC 985} &$(-8,2)$ &2 &$-8,2$ &$-7$ &-- &--\\
{NGC 1068} &$-58,-6,(5)$ &5 &$-6,5$ &16 &$-58$,None &1\\
{NGC 1705} &-- &-- &-- &-- &87,None &--\\
{NGC 4151} &$(-41,-29),-24,-21,0$ &$-41$ &$-41,-29,-24,-21$ &$-35$ &$-29$,Yes &3\\
{NGC 4670} &$-28,(-19,-8)$ &$-8$ &$-19,-8$ &$-11$ &-- &--\\
{NGC 7469} &$-13,(-3,0)$ &0 &$-13,-3,0$ &$-1$ &-- &--\\
{PG 0804+761} &$-58,(-10,-5,6)$ &$-5$ &$-58,-10,-5,6$ &4 &$-58$,Yes &2\\
{PG 0844+349} &$-12,(1,6)$ &6 &1,6 &8 &-- &--\\
{PG 0953+414} &$-49,(-4)$ &$-4$ &$-4$ &3 &$-49$,None &1\\
{PG 1116+215} &-- &-- &-- &-- &$-42$,Yes &--\\
{PG 1211+143} &$-35,(-20,-9)$ &$-20$ &$-35,-20,-9$ &$-12$ &$-35$,None &2\\
{PG 1259+593} &$-54,(-6)$ &$-6$ &$-6$ &$-5$ &$-54$,Yes &1\\
{PG 1302$-$102} &$-53,(-22,-3,-1),14$ &$-1$ &$-22,-3,-1,14$ &$-8$ &-- &--\\
{PKS 0405$-$12} &$(-13,1,4),18$ &1 &$-13,1,4,18$ &0 &-- &--\\
{PKS 0558$-$504} &(4,6),26,27,71 &6 &4,6,26,27 &10 &26,None &2\\
{} & & & & &71,None &1\\
{PKS 2005$-$489} &$-26,(2)$ &2 &2 &$-1$ &-- &--\\
{PKS 2155$-$304} &$(-4),14$ &$-4$ &$-4$ &$-8$ &-- &--\\
{Ton S180} &-- &-- &-- &-- &-- &--\\
{Ton S210} &$-30,(-10,-6)$ &$-6$ &$-30,-10,-6$ &$-10$ &-- &--\\
{VII Zw 118} &$(-7,-3,7)$ &7 &$-7,-3,7$ &$-2$ &-- &--\\
\enddata
\tablenotetext{a}{$v$(H~I) lists the 21~cm velocity components fitted by Wakker 
\etal\ (2003), excluding high-velocity components at $|v| > 90$ \kms.  The components in 
parentheses are those we have selected as most likely to correspond to the \Htwo\ absorption, 
based on the coincidence of radial velocities (\S~2.3).  The sum of the column densities
for the components in parentheses is the value we adopt for \NHI.}
\tablenotetext{b}{$v_{\rm min}$(H~I) indicates the velocity component with the mimimum column
density that, based on the coincidence of radial velocities, could be solely 
associated with the \Htwo.  The column density of this component provides the lower bound
for \NHI.}
\tablenotetext{c}{$v_{\rm max}$(H~I) lists all velocity components that, based on the
coincidence of radial velocities, could feasibly be associated with the \Htwo.  The sum
of the column densities of these components provides the upper bound for \NHI.}
\tablenotetext{d}{In general, the uncertainty on $v$(\Htwo) is $\sim\pm{5}$ \kms\ for 
$\rm{(S/N)_{bin}>4}$.}
\tablenotetext{e}{Velocity and \Htwo\ status of IVCs presented in Table 2 of 
   Richter \etal\ (2003).}
\tablenotetext{f}{Notes on IVCs:  
(1) IVC does not contribute to fitted \Htwo\ lines;   
(2) Possible IVC contribution on edge of fitted \Htwo\ lines; 
(3) IVC is main fitted \Htwo\ component.  
Although we generally did not measure \Htwo\ in IVCs, we did so for  
targets noted as (3).  For 3C 273, we do not consider component at 25 \kms\ 
to be true IVC. For Mrk 876, we fitted the IVC lines blended with LSR
lines (double component fits). For NGC 4151, the \Htwo\ lines may have 
some contributions from non-IVC gas.}
\tablenotetext{g}{Since this sight line has two H~I components at the same velocity, 
$-5$ \kms, we distinguish them by ``b" for broad and ``n" for narrow.}
\end{deluxetable}

\clearpage


\begin{deluxetable}{lcccccc}
\tabletypesize{\scriptsize}
\tablecolumns{7}
\tablenum{4}
\tablewidth{0pt}
\tablecaption{\Htwo\ SIGHTLINE SUMMARY}

\tablehead{
\colhead{Name}
&\colhead{log~\NHtwo}
&\colhead{log~\NHI}
&\colhead{beam\tablenotemark{a}}
&\colhead{$\rm{log~f_{H2}}$}
&\colhead{$T_{\rm{01}}$}
&\colhead{$T_{\rm{exc}}$}\\
\colhead{}
&\colhead{(\cd)}
&\colhead{(\cd)}
&\colhead{size}
&\colhead{}
&\colhead{(K)}
&\colhead{(K)}
}

\startdata
{3C 249.1} &18.98 $\pm^{0.14}_{0.16}$ &20.25 $\pm^{0.20}_{0.44}$ &3 &$-1.01$ &
   144 $\pm^{184}_{47}$ &406 $\pm{317}$\\
{3C 273 ($v=-5$)} &14.30 $\pm^{0.05}_{0.05}$ &20.14 $\pm^{0.01}_{0.37}$ &3 &
   $-5.54$ &-- & --\\
{\hspace{9.3mm} ($v=26$)} &15.72 $\pm^{0.24}_{0.13}$ &19.42 $\pm^{0.03}_{0.03}$ &
   3 &$-3.4 0$ &216 $\pm^{624}_{77}$ &645 $\pm{71}$\\
{ESO 141$-$G55} &19.32 $\pm^{0.07}_{0.07}$ &20.70 $\pm^{0.08}_{0.67}$ &2 &$-1.12$ &98 
   $\pm^{22}_{15}$ &416 $\pm{72}$\\
{H 1821+643} &17.91 $\pm^{0.13}_{0.20}$ &20.43 $\pm^{0.14}_{0.57}$ &4 &$-2.22$ &-- &494 $
\pm{168}$\\
{HE 0226$-$4110}\tablenotemark{b} &$\le$14.29 &$\ge$19.50 &2 &$\le$$-4.92$ &-- &--\\
{HE 1143$-$1810} &16.54 $\pm^{1.32}_{0.68}$ &20.47 $\pm^{0.03}_{1.19}$ &3 &$-3.63$ &150 $
\pm^{206}_{74}$ &484 $\pm{185}$\\
{HS 0624+6907} &19.82 $\pm^{0.10}_{0.10}$ &20.80 $\pm^{0.09}_{1.06}$ &4 &$-0.76$ &100 
  $\pm^{34}_{18}$ &498 $\pm{34}$\\
{MRC 2251$-$178} &14.54 $\pm^{0.23}_{0.17}$ &20.39 $\pm^{0.01}_{1.55}$ &3 &$-5.55$ &-- &-
-\\
{Mrk 9} &19.36 $\pm^{0.09}_{0.08}$ &20.64 $\pm^{0.04}_{1.14}$ &3 &$-1.02$ &115 $\pm^{48}_
{25}$ &215 $\pm{37}$\\
{Mrk 106} &16.23 $\pm^{0.21}_{0.15}$ &20.35 $\pm^{0.03}_{0.83}$ &4 &$-3.82$ &68 $\pm^{21}
_{12}$ &578 $\pm{123}$\\
{Mrk 116} &19.08 $\pm^{0.13}_{0.13}$ &20.41 $\pm^{0.07}_{0.42}$ &4 &$-1.06$ &71 $\pm^{16}
_{12}$ &437 $\pm{82}$\\
{Mrk 205} &16.53 $\pm^{0.13}_{0.37}$ &20.40 $\pm^{0.03}_{0.39}$ &4 &$-3.57$ &-- &762 $\pm
{542}$\\
{Mrk 209} &$\le$14.48 &$\ge$19.73 &3 &$\le$$-4.95$ &-- &--\\
{Mrk 290} &16.18 $\pm^{2.01}_{0.39}$ &20.11 $\pm^{0.03}_{0.93}$ &4 &$-3.62$ &76 $\pm^{54}
_{26}$ &592 $\pm{260}$\\
{Mrk 335} &18.83 $\pm^{0.08}_{0.08}$ &20.43 $\pm^{0.14}_{1.30}$ &3 &$-1.32$ &92 $\pm^{27}
_{14}$ &434 $\pm{56}$\\
{Mrk 421} &14.63 $\pm^{0.09}_{0.10}$ &19.94 $\pm^{0.03}_{1.16}$ &3 &$-5.00$ &167 $\pm^{11
1}_{105}$ &--\\
{Mrk 478} &$\le$14.56 &$\ge$19.21 &3 &$\le$$-4.35$ &-- &--\\
{Mrk 501} &14.78 $\pm^{0.18}_{0.10}$ &20.24 $\pm^{0.03}_{1.01}$ &4 &$-5.16$ &-- &--\\
{Mrk 509} &17.87 $\pm^{0.31}_{0.78}$ &20.58 $\pm^{0.03}_{0.58}$ &3 &$-2.41$ &
    123 $\pm^{262}_{40}$ &371 $\pm{180}$\\
{Mrk 817} &$\le$14.03 &$\ge$19.83 &4 &$\le$$-5.50$ &-- &--\\
{Mrk 876 ($v=-33$)} &15.75 $\pm^{2.19}_{0.32}$ &19.84 $\pm^{0.03}_{0.03}$ &4 &$-3.79$ &
   123 $\pm^{189}_{76}$ &689 $\pm{339}$\\
{\hspace{11mm} ($v=-3$)} &16.58 $\pm^{1.96}_{0.42}$ &20.21 $\pm^{0.03}_{0.03}$ &4 &
   $-3.33$ &-- &509 $\pm{248}$\\
{Mrk 1095} &18.76 $\pm^{0.21}_{0.31}$ &20.95 $\pm^{0.02}_{1.04}$ &1 &$-1.90$ &
   121 $\pm^{107}_{39}$ &501 $\pm{129}$\\
{Mrk 1383} &14.35 $\pm^{0.13}_{0.10}$ &20.40 $\pm^{0.03}_{0.34}$ &3 &$-5.75$ &-- &--\\
{Mrk 1513} &16.42 $\pm^{1.08}_{0.26}$ &20.52 $\pm^{0.03}_{0.99}$ &3 &$-3.81$ &70 $\pm^{30}_{21}$ &514 $\pm{156}$\\
{MS 0700.7+6338} &18.75 $\pm^{0.27}_{0.68}$ &20.43 $\pm^{0.18}_{0.50}$ &1 &$-1.40$ &82 $\pm^{61}_{21}$ &200 $\pm{76}$\\
{NGC 985} &16.05 $\pm^{1.95}_{0.33}$ &20.52 $\pm^{0.03}_{1.16}$ &3 &$-4.17$ &102 $\pm^{76}_{32}$ &572 $\pm{183}$\\
{NGC 1068} &18.13 $\pm^{0.13}_{0.17}$ &19.61 $\pm^{0.81}_{0.03}$ &3 &$-1.20$ &76 $\pm^{23}_{14}$ &471 $\pm{209}$\\
{NGC 1705} &$\le$14.17 &$\ge$19.66 &2 &$\le$$-5.20$ &-- &--\\
{NGC 4151} &16.70 $\pm^{0.93}_{0.31}$ &20.20 $\pm^{0.10}_{0.95}$ &3 &$-3.20$ &-- &504 $\pm{82}$\\
{NGC 4670} &14.72 $\pm^{0.13}_{0.16}$ &19.95 $\pm^{0.03}_{0.48}$ &1 &$-4.93$ &98 $\pm^{121}_{115}$ &--\\
{NGC 7469} &19.67 $\pm^{0.10}_{0.10}$ &20.59 $\pm^{0.05}_{1.46}$ &4 &$-0.71$ &71 $\pm^{16}_{11}$ &389 $\pm{35}$\\
{PG 0804+761} &18.66 $\pm^{0.14}_{0.19}$ &20.54 $\pm^{0.04}_{1.00}$ &4 &$-1.60$ &144 $\pm^{162}_{44}$ &363 $\pm{105}$\\
{PG 0844+349} &18.22 $\pm^{0.18}_{0.28}$ &20.34 $\pm^{0.03}_{0.68}$ &3 &$-1.82$ &147 $\pm^{312}_{50}$ &311 $\pm{87}$\\
{PG 0953+414} &15.03 $\pm^{0.11}_{0.10}$ &20.00 $\pm^{0.03}_{0.03}$ &4 &$-4.67$ &252 $\pm^{124}_{108}$ &--\\
{PG 1116+215} &$\le$14.16 &$\ge$19.70 &4 &$\le$$-5.30$ &-- &--\\
{PG 1211+143} &18.38 $\pm^{0.15}_{0.14}$ &20.25 $\pm^{0.17}_{1.03}$ &3 &$-1.58$ &142 $\pm^{90}_{38}$ &321 $\pm{127}$\\
{PG 1259+593} &14.75 $\pm^{0.10}_{0.12}$ &19.67 $\pm^{0.03}_{0.03}$ &4 &$-4.62$ &193 $\pm^{115}_{108}$ &--\\
{PG 1302$-$102} &15.62 $\pm^{1.41}_{0.16}$ &20.42 $\pm^{0.08}_{1.26}$ &3 &$-4.50$ &-- &671 $\pm{233}$\\
{PKS 0405$-$12} &15.79 $\pm^{0.25}_{0.12}$ &20.41 $\pm^{0.13}_{0.89}$ &3 &$-4.32$ &-- &852 $\pm{184}$\\
{PKS 0558$-$504} &15.44 $\pm^{0.18}_{0.12}$ &20.53 $\pm^{0.11}_{0.46}$ &2 &$-4.78$ &-- &671 $\pm{173}$\\
{PKS 2005$-$489} &15.07 $\pm^{0.10}_{0.10}$ &20.60 $\pm^{0.03}_{0.03}$ &2 &$-5.23$ &139 $\pm^{121}_{105}$ &729 $\pm{372}$\\
{PKS 2155$-$304} &14.17 $\pm^{0.11}_{0.14}$ &20.06 $\pm^{0.03}_{0.03}$ &4 &$-5.59$ &147 $\pm^{115}_{111}$ &--\\
{Ton S180} &$\le$14.37 &$\ge$20.08 &3 &$\le$$-5.41$ &-- &--\\
{Ton S210} &16.57 $\pm^{1.18}_{1.38}$ &20.19 $\pm^{0.01}_{1.11}$ &4 &$-3.32$ &144 $\pm^{192}_{72}$ &--\\
{VII Zw 118} &18.84 $\pm^{0.10}_{0.12}$ &20.56 $\pm^{0.03}_{0.80}$ &3 &$-1.43$ &108 $\pm^{48}_{24}$ &552 $\pm{80}$\\
\enddata
\tablenotetext{a}{Source of 21~cm data: (1) Leiden-Dwingeloo Survey ($35'$ beam), 
(2) Villa Elisa telescope ($34'$ beam), (3) Green Bank 140-ft telescope
($21'$ beam), (4) Effelsberg telescope ($9'.7$ beam).}
\tablenotetext{b}{Savage \etal\ (2005) have detected \Htwo\ in this sightline,
with log~\NHtwo\ $\approx 14.39 \pm 0.43$.} 

\end{deluxetable}

\clearpage


\begin{figure}[!htb]
\epsscale{1.0}
\plotone{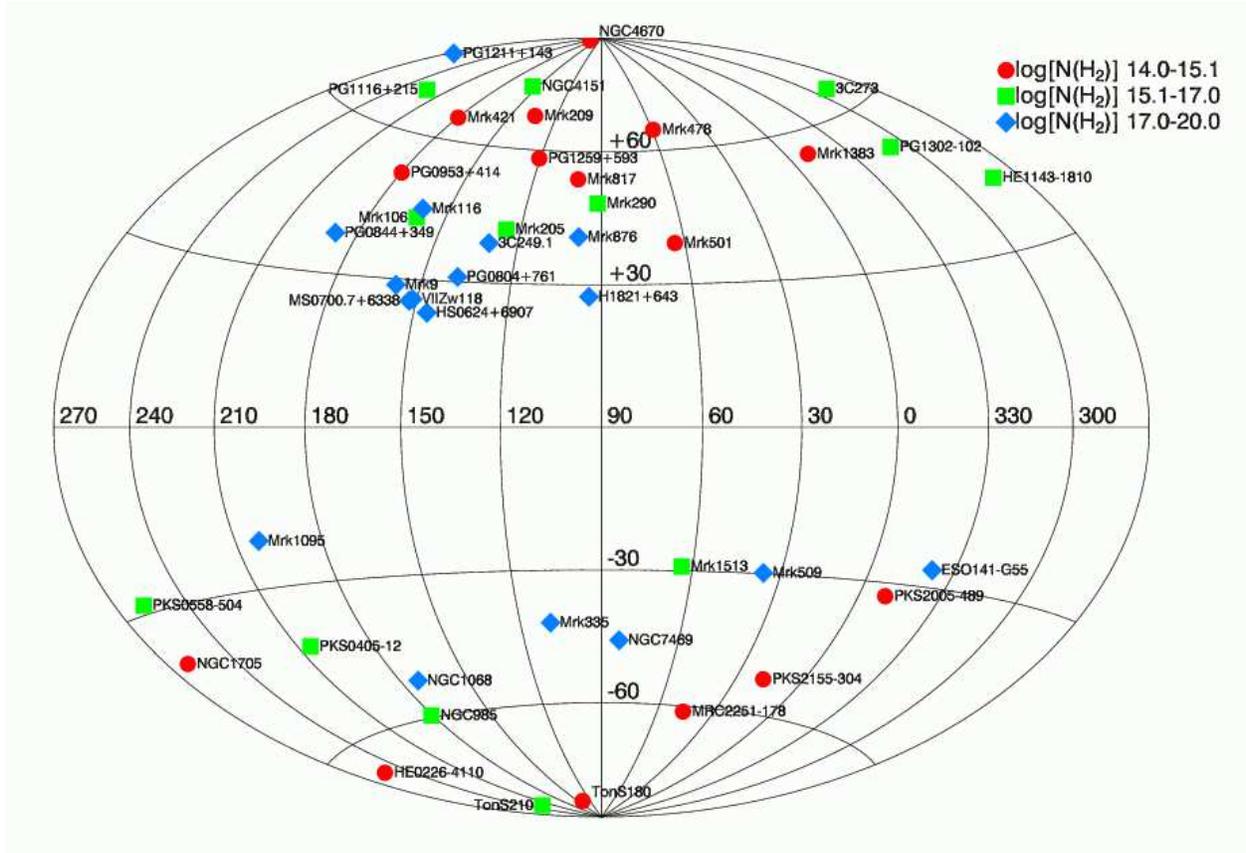}
\caption{Aitoff projection showing locations of 45 AGN targets from
our high-latitude \Htwo\ survey. In the northern Galactic hemisphere,
we identified 8 high-latitude ($b > 54^{\circ}$) sight lines with low
\NHtwo\ $\leq 10^{15.1}$ \cd, color-coded in red.
These targets suggest a widespread
\Htwo\ hole ($60^{\circ} < \ell < 180^{\circ}$ and $b > 54^{\circ}$).
Sightlines with higher \NHtwo\ are labeled in green and blue.
We sampled targets with Galactic latitude $|b| > 20^{\circ}$ to
minimize absorption by interstellar gas in the Galactic disk.
The low column density ``H$_{2}$ Holes" at $|b| > 54^{\circ}$ may be
related to the ``Northern Chimney" (region of low Na~I absorption,
Lallement et al.\ 2003) and the ``Lockman Hole" (region near
of low \NHI\ near $\ell = 135^{\circ}-175^{\circ}$ and
$b = 46^{\circ}-60^{\circ}$, Lockman \etal\ 1986, see also
Figure 10). These ``\Htwo-clean" sight lines are ideal for FUV studies
of the IGM and HVCs.}
\label{aitoff}
\end{figure}

\clearpage


\begin{figure}[!htb]
\epsscale{0.75}
\plotone{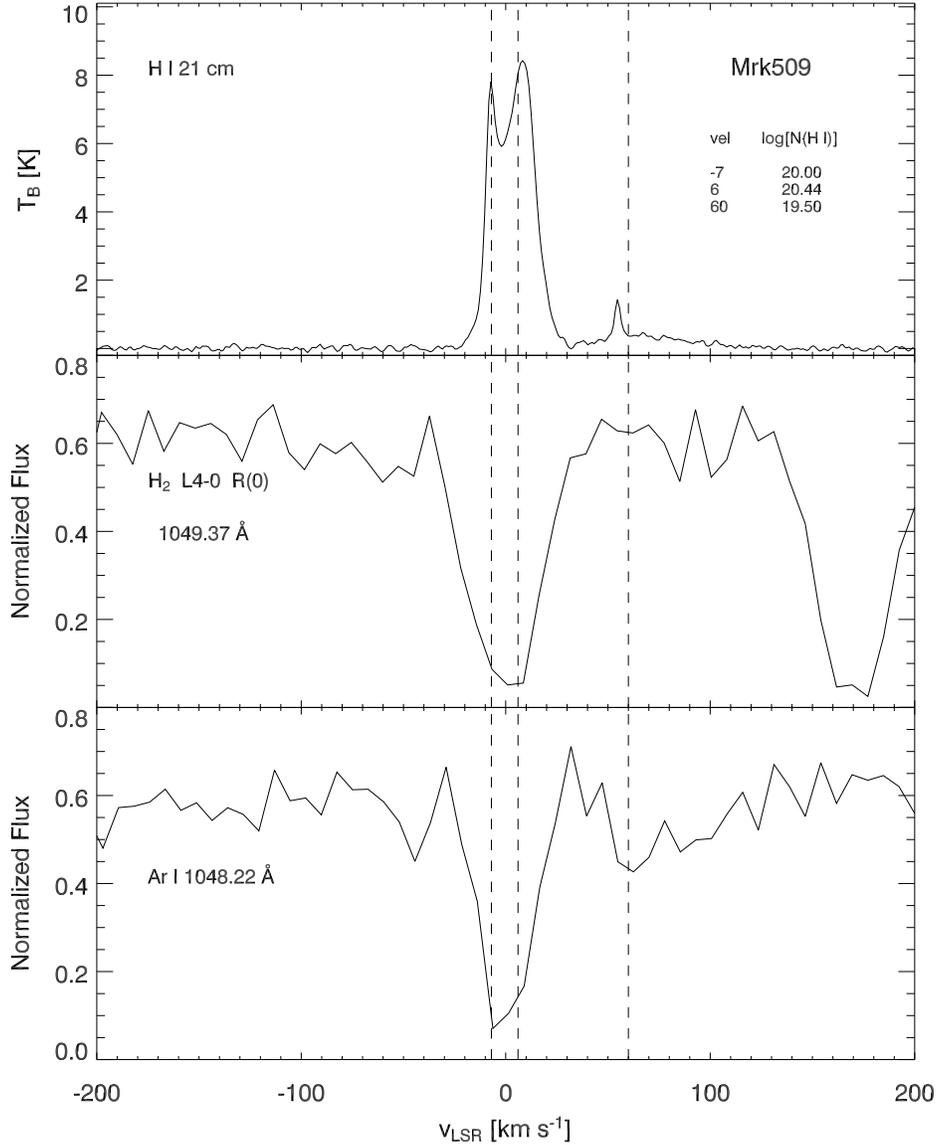}
\caption{{\it Top Panel}: 21 cm spectrum of Mrk~509 from Leiden-Dwingeloo
Survey. {\it Middle and bottom panels}: \FUSE\ spectra binned by 4 pixels.
Dotted lines indicate the three \ion{H}{1} components identified by Wakker
\etal\ (2003) listed in top panel.  \FUSE\ data have been shifted
to align the \ion{H}{1} emission and \ion{Ar}{1} absorption.  A comparison
of \ion{H}{1} and \Htwo\ shows that either the $-7$ \kms\
H~I component or the 6 \kms\ H~I component, or both, could be associated with 
the \Htwo\ absorption.  We adopt the sum of the two column densities, 
log~\NHtwo\ = 20.58, for \NHI.  Since either component could be solely 
responsible for the \Htwo\ absorption, we choose the lower column density, 
log~\NHtwo\ =20.00, for our lower limit on \NHI.  Richter \etal\ (2003) detect 
\Htwo\ in the 60 \kms\ component, but this absorption does not contribute to 
our fitted \Htwo\ lines.  Since no other H~I components can contribute to the 
\Htwo\ absorption, we adopt the sum of the $-7$ \kms\ component and 6 \kms\ 
component (plus a systematic uncertainty, $\Delta$ log~\NHI\ = 0.03) for the 
upper limit on \NHI.}
\label{vp_mrk509}
\end{figure}

\clearpage


\begin{figure}[!htb]
\epsscale{0.75}
\plotone{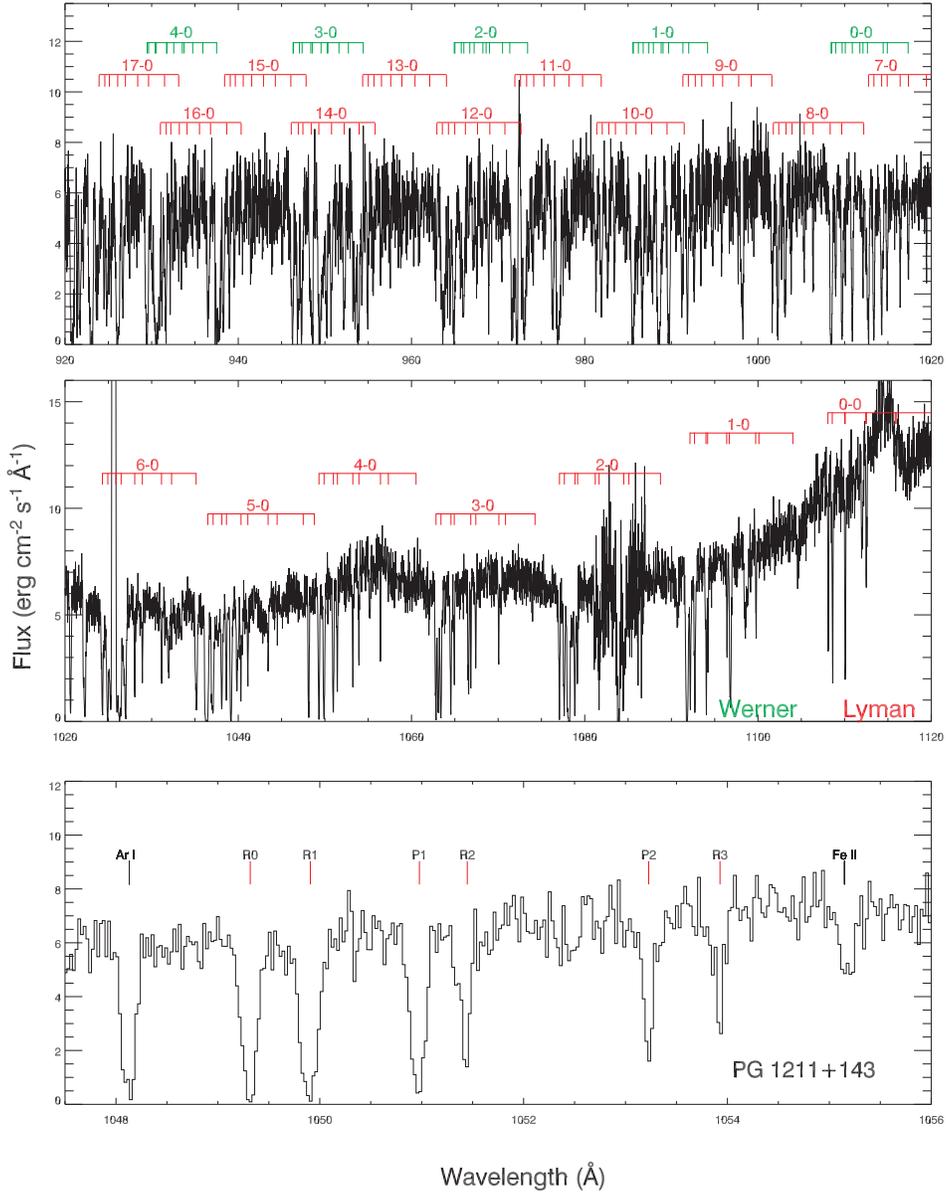}
\caption{ {\it Upper two panels}: \FUSE\ spectrum of PG~1211+143,
 showing expected locations of (0-0) through (17-0) H$_2$ Lyman
 bands (red) and (0-0) through (4-0) Werner bands (green).
 This sight line has log~\NHtwo\ = $18.38^{+0.15}_{-0.14}$
 in $J=0-3$.
 {\it Lower panel:} Blowup of the (4-0) Lyman band of H$_2$,
 showing R(0), R(1), P(1), R(2), P(2), and R(3) absorption lines
 from $J$ = 0, 1, 2, and 3 together with interstellar lines of
 Ar~I $\lambda 1048.220$ and Fe~II $\lambda 1055.262$. }
\label{PG1211}
\end{figure}

\clearpage

i

\begin{figure}[!htb]
\epsscale{0.7}
\plotone{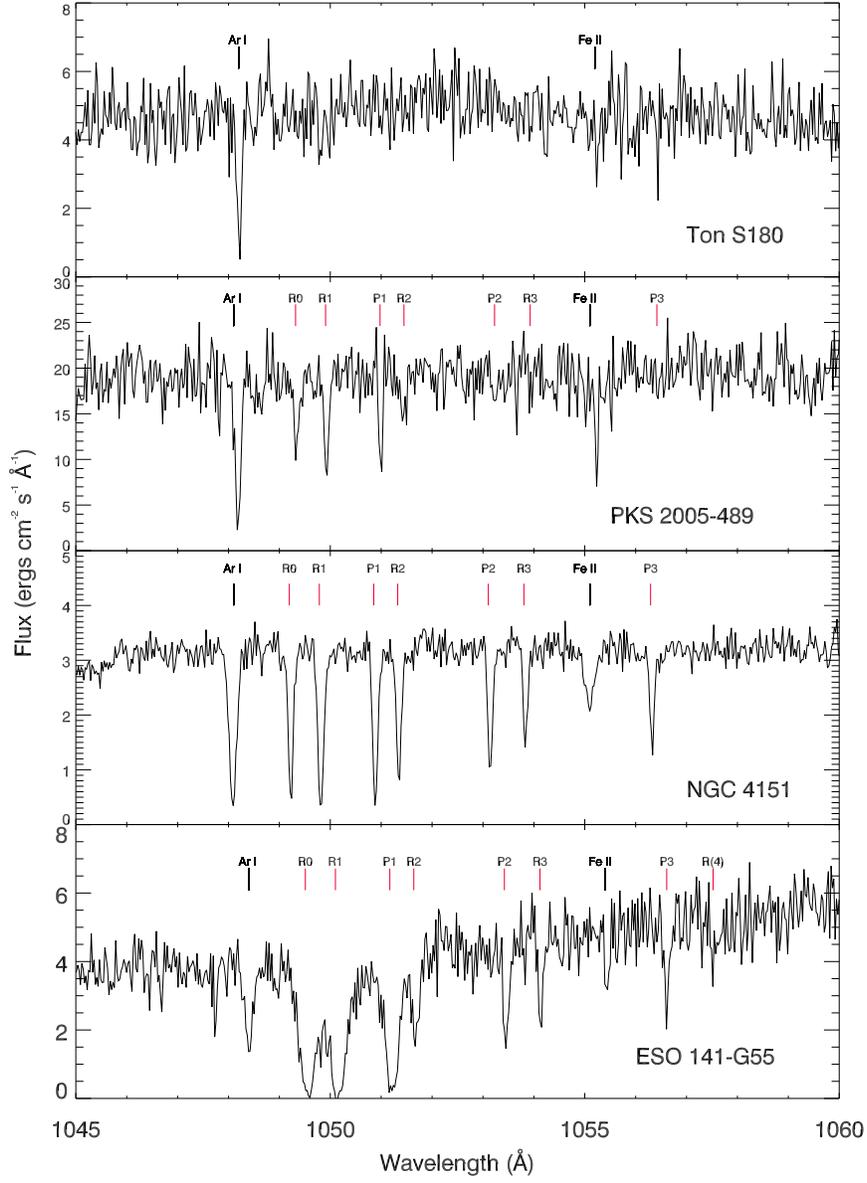}
\caption{Examples of \FUSE\ spectra of four AGN, showing increasing
  H$_2$ column densities from the Lyman (4--0) band.
  {\it First panel:}  Ton~S180 shows no detectable to a limit
  log \NHtwo\ $<14.37$ ($<13.98$ in $J=0$ and $<14.14$ in $J=1$).
  {\it Second panel:}  PKS~2005-489 with log \NHtwo\ =
  $15.07^{+0.10}_{-0.10}$.  {\it Third panel:} NGC~4151 with
  log \NHtwo\ = $16.70^{+0.93}_{-0.31}$.
  {\it Fourth panel:} ESO~141-G55 with log \NHtwo\ =
  $19.32^{+0.07}_{-0.07}$. }
\label{Nmontage}
\end{figure}

\clearpage


\begin{figure}[!htb]
\epsscale{1.0}
\plotone{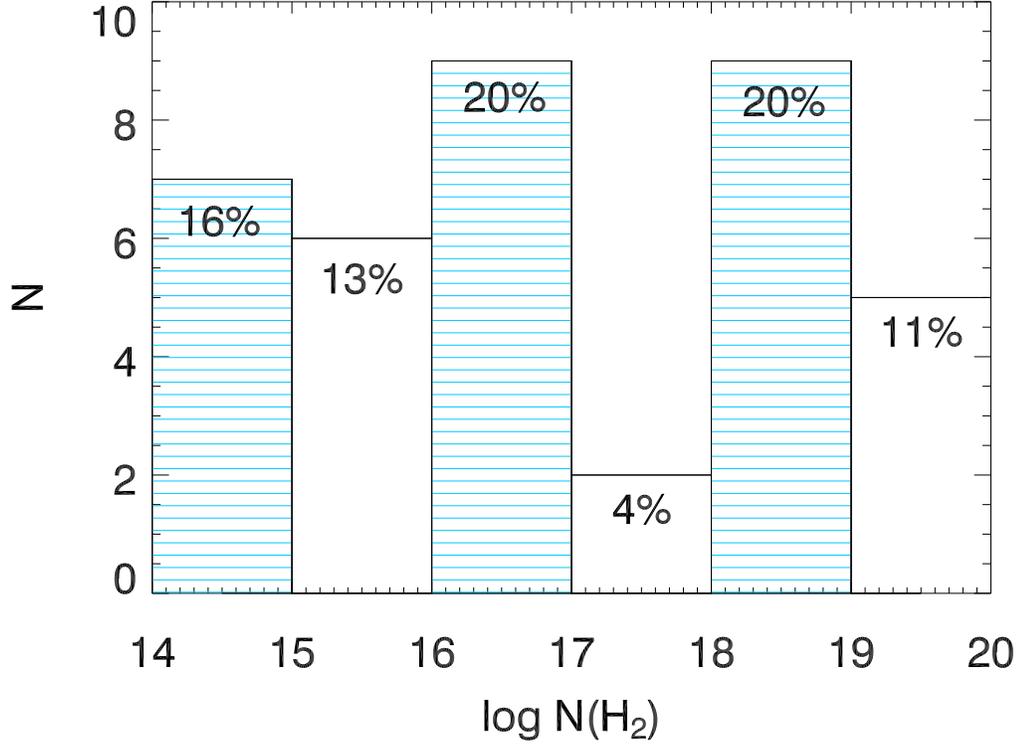}
\caption{Distribution of \NHtwo\ (the LSR absorption only) for 38 of 
our 45 AGN sight lines.  An additional 7 sight lines (16\% of the sample) 
showed no detectable low-velocity H$_{2}$ above a threshold, 
log~\NHtwo\ = 13.8--14.6, depending on the S/N (2-11 per pixel) and 
resolution ($R=15,000-20,000$ for {\it FUSE}). The dip in numbers between 
log~\NHtwo\ = 17 -- 18 may divide two populations (see also Fig. 9), 
although this column density bin is sensitive to CoG uncertainties and 
is based on small numbers of sight lines. }
\label{Ndistrib}
\end{figure}

\clearpage


\begin{figure}[!htb]
\epsscale{1.0}
\plotone{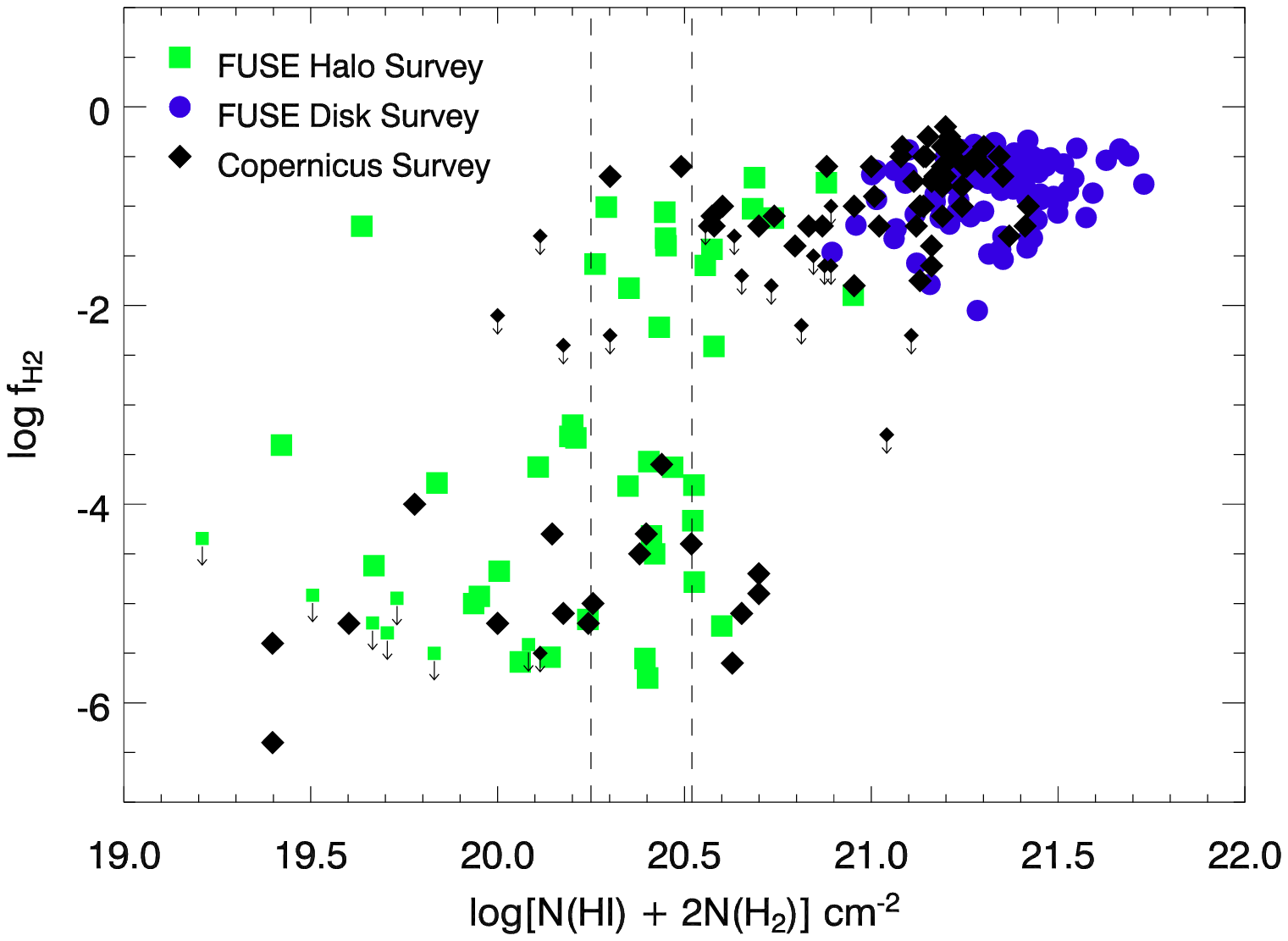}
\caption{Transitions in molecular fraction, $f_{\rm H2}$, with
total hydrogen column density, N$_{\rm H}$ = (\NHI\ + 2\NHtwo ). In
our \FUSE\ survey of 45 high-latitude sight lines, the transition occurs
at log N$_{\rm H}  \approx 20.38 \pm 0.13$, approximately a factor of two
below the transition (log N$_{\rm H}$ = 20.7) in the {\it Copernicus}
survey (Savage \etal\ 1977). This shift to lower N$_H$ could arise
from reduced photodissociation (lower UV field), from enhanced
\Htwo\ formation rates (higher $n_H$ in compressed cirrus clouds) or both. }
\label{fH2}
\end{figure}

\clearpage


\begin{figure}[!htb]
\epsscale{1.0}
\plotone{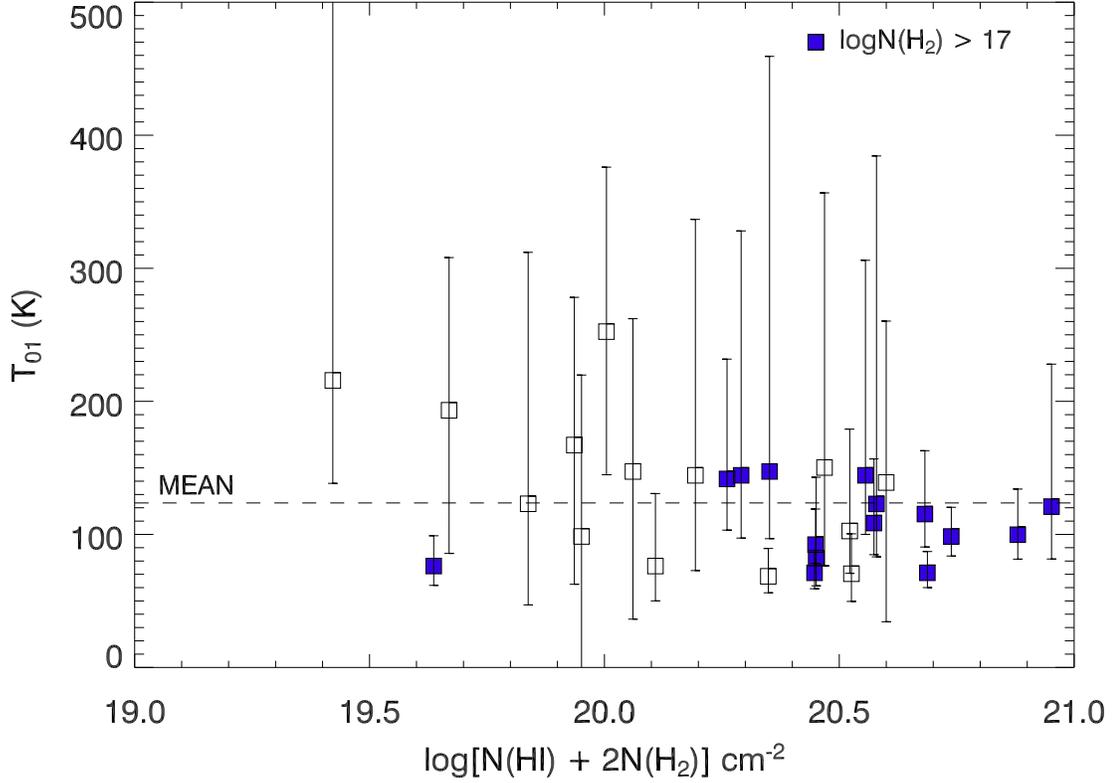}
\caption{Mean rotational temperature in our Galactic high-latitude survey
is $\langle T_{01} \rangle = 124\pm8$~K (median 121~K). For 15 sight lines 
with log~\NHtwo\ $\geq 17$ (solid squares), we find 
$\langle T_{01} \rangle = 109\pm7$~K. These temperatures are somewhat higher 
than those found in the {\it Copernicus} \Htwo\ survey ($77 \pm 17$~K, 
Savage \etal\ 1977) and in our \FUSE\ Galactic disk survey ($86 \pm 20$~K, 
Shull \etal\ 2005). 
The range of $T_{01}$ is 68--252~K, consistently higher than the
Galactic disk values, all but two of which lie between 55--120~K.}
\label{T01}
\end{figure}

\clearpage


\begin{figure}[!htb]
\epsscale{1.0}
\plotone{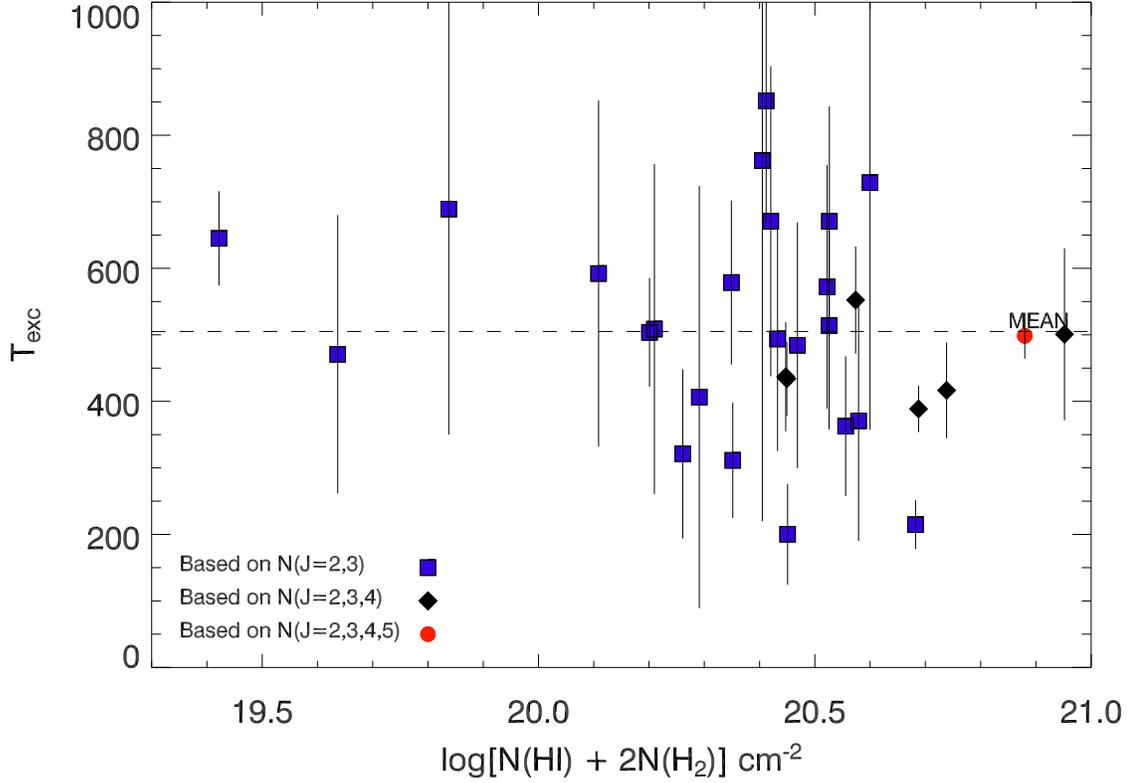}
\caption{
The mean excitation temperature for $J \geq 2$ is
$\langle T_{\rm exc}\rangle=505\pm{28}$~K (median 501~K), based
on 30 sight lines with detectable $J=3$ lines (in some cases $J=4$ and
$J=5$).  This value is slightly larger than the mean, $326 \pm 125$~K,
seen in the \FUSE\ disk \Htwo\ survey (Shull \etal\ 2005). The
distribution of $T_{\rm exc}$ is a
measure of non-thermal excitation by UV pumping and perhaps H$_2$
formation on grain surfaces (Browning \etal\ 2003).}
\label{Texc}
\end{figure}

\clearpage


\begin{figure}[!htb]
\epsscale{0.75}
\plotone{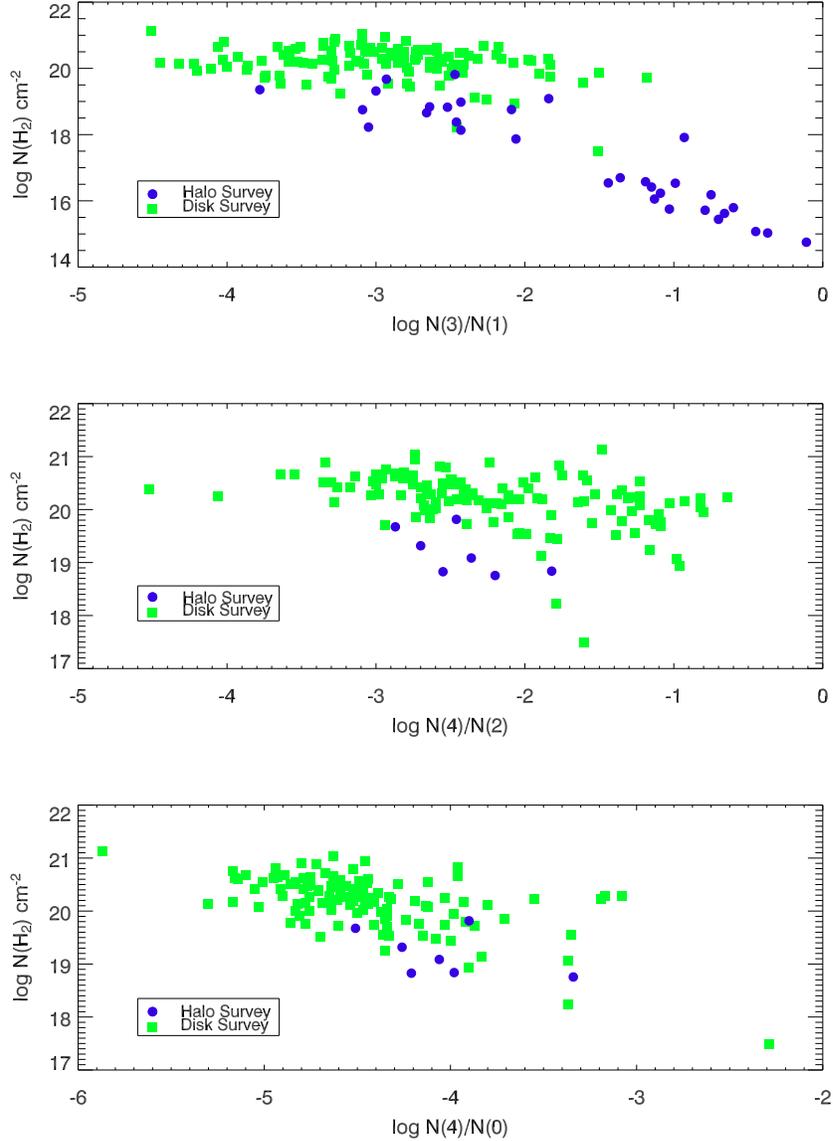}
\caption{Ratios of rotational-state column densities for $J = 3/1$,
$J = 4/2$, $J = 4/0$ vs. \NHtwo\ for sight lines in our surveys in
the Galactic halo (blue) and Galactic disk (green).  Of the
45 halo sight lines, 8 have measured column densities for both
$J = 3$ and $J=4$, while 32 have $J=3$ but lack detectable $J=4$
because of insufficient total \NHtwo.  For all sight lines with
log~\NHtwo\ $\geq 18$, the degree of rotational excitation
is comparable (Shull \etal\ 2005), except for a population of 16 halo 
sight lines with {\it low} \NHtwo\ and highly excited $J=3$ 
[log~N(3)/N(1) $\geq -1.5$].
A comparison of the high-\NHtwo\ data with models of \Htwo\ formation
and rotational excitation (Browning \etal\ 2003) suggests that the
UV radiation fields in the low Galactic halo and disk are similar.
To produce the observed shift to lower N$_{\rm H}$ in the transition of
$f_{\rm H2}$, the \Htwo\ formation rate may be enhanced in the halo
absorbers, because of higher $n_H$ in compressed cirrus clouds. }
\label{J-ratios}
\end{figure}

\clearpage


\begin{figure}[!htb]
\epsscale{1.0}
\plotone{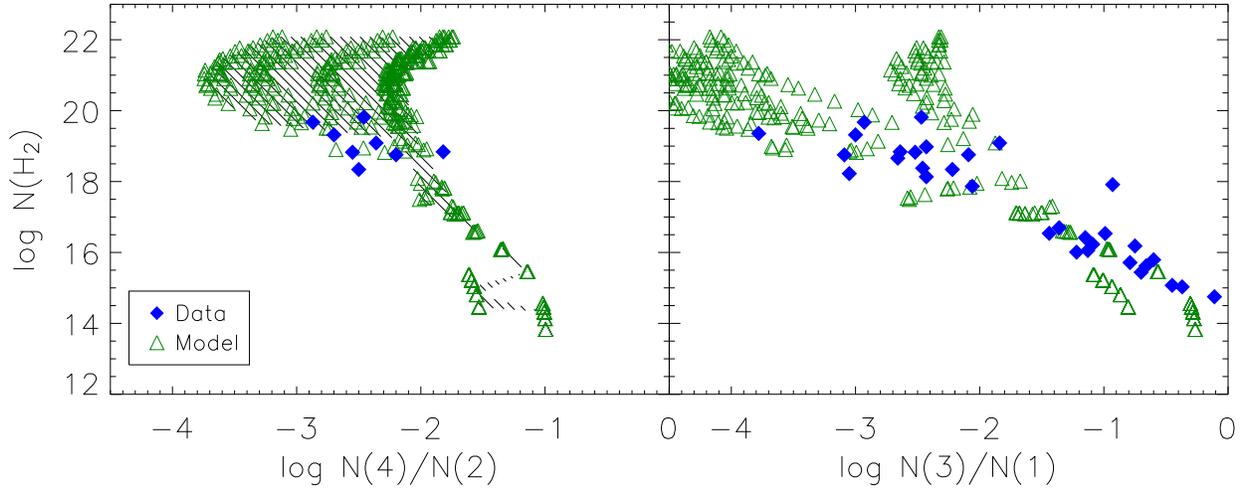}
\caption{
Models of \Htwo\ rotational excitation ratios, $N(J)$, using the code 
of Browning, Tumlinson, \& Shull (2003).  Plots show ratios,
$N(4)/N(2)$ and $N(3)/N(1)$, versus \NHtwo, for data (blue points, 
same as in Fig.\ 9) and for a grid of models (green points).  
Models assume the mean Galactic \Htwo\ formation rate  
($R = 3 \times 10^{-17}$ cm$^3$~s$^{-1}$), and a range of parameters, 
FUV radiation fields, varying by a factor of two above and below 
the mean ISM value ($I_{\rm FUV} = 2 \times 10^{-8}$ photons cm$^{-2}$
s$^{-1}$ Hz$^{-1}$ between 912--1120 \AA).   
The model grid ponts include variations in cloud temperatures
($T=$ 20--150~K) and density ($n_H$ = 5--800~cm$^{-3}$).  
Note the good agreement between data and models, which show increasing
excitation, $N(3)/N(1)$, for ``thinner" clouds, log~\NHtwo\ $\leq 18$. }  
\label{models}
\end{figure}

\clearpage


\begin{figure}[!htb]
\epsscale{1.0}
\plotone{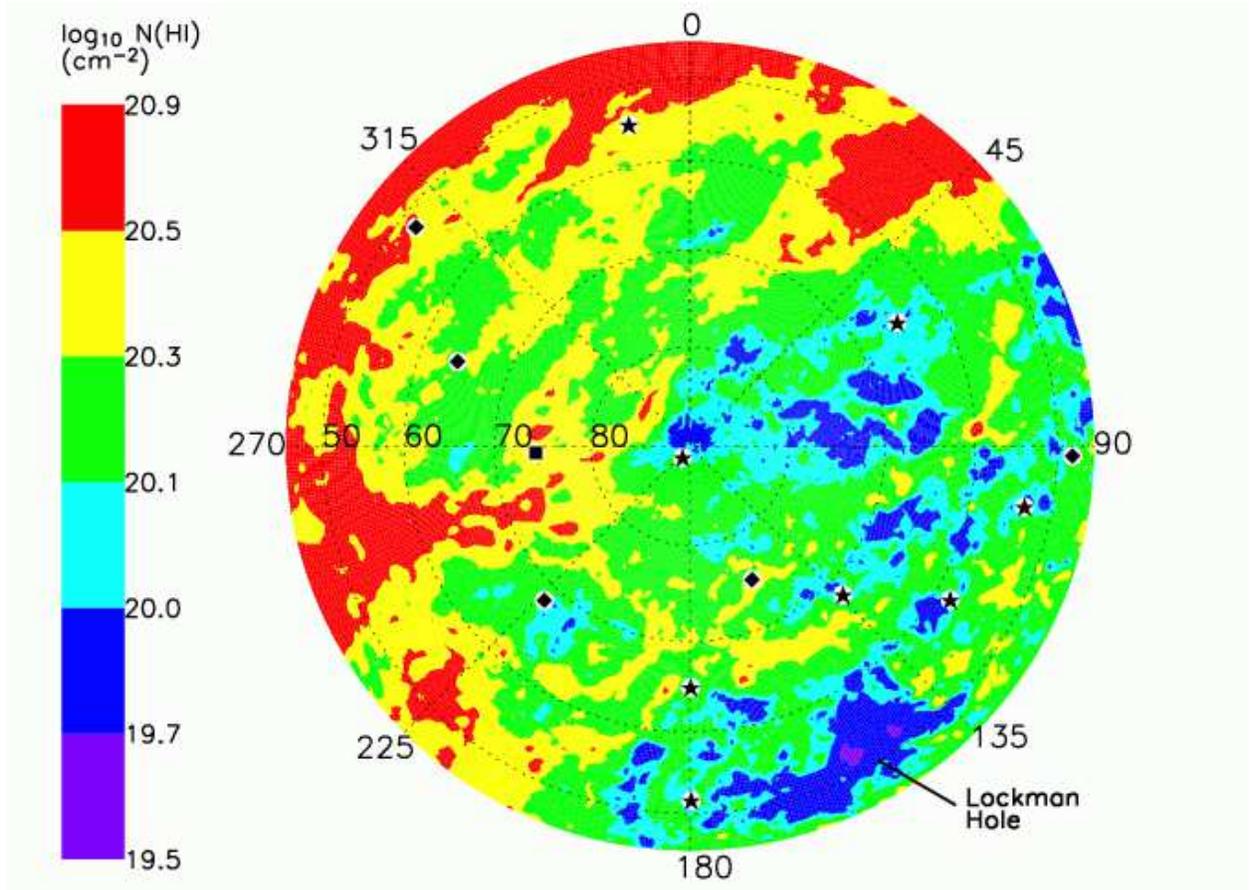}
\caption{
Locations in northern Galactic hemisphere ($b > 45^{\circ}$) of 14
high-latitude AGN sight lines (white circles). The symbols inside
white circles indicate log~\NHtwo\ in the ranges:
14.0--15.1 (Star); 15.1--17.0 (Diamond); 17.0--20.0 (Square).
The targets sample the distribution of \NHI\ in the
Leiden Dwingeloo Survey (Hartmann \& Burton 1997), colored as
deep-blue (log~\NHI\ = 19.7--20.0) or light-blue (20.0--20.1).
Gas at higher \NHI\ is shown in green, yellow, or red. The
extended Lockman Hole (log~\NHI\ $\approx$ 19.5--20.0) is seen
in blue/purple ($\ell = 135^{\circ}-175^{\circ}$ and
$b = 46^{\circ}-60^{\circ}$) at lower right. }
\label{polar}
\end{figure}


\begin{thebibliography}{}

\bibitem[Abgrall et al. 1993a]{Abg93a}
Abgrall, H., Roueff, E., Launay, F., Roncin, J. Y.,
\& Subtil, J. L. 1993a, \aaps, 101, 273

\bibitem[Abgrall et al. 1993a]{Abg93b}
Abgrall, H., Roueff, E., Launay, F., Roncin, J. Y.,
\& Subtil, J. L. 1993b, \aaps, 101, 323

\bibitem[Black \& Dalgarno 1973]{Bl73}
Black, J. H., \& Dalgarno, A. 1973, \apj, 184, L101

\bibitem[Black \& Dalgarno 1976]{Bl76}
Black, J. H., \& Dalgarno, A. 1976, \apj, 203, 132

\bibitem[Bland-Hawthorn \& Maloney 1999]{Bland99}
Bland-Hawthorn, J., \& Maloney, P. R. 1999, \apj, 510, L33  

\bibitem[Blitz et al. 1984]{Blitz84}
Blitz, L., Magnani, L., \& Mundy, L. 1984, \apj, 282, L9

\bibitem[Browning, Tumlinson, Shull (2002)]{BTS03}
Browning, M. K., Tumlinson, J., \& Shull, J. M. 2003,
\apj, 582, 810

\bibitem[Collins et al. (2003)]{Co03}
Collins, J. A., Shull, J. M., \& Giroux, M. L. 2003,
\apj, 585, 336

\bibitem[Collins et al. (2004)]{Co04}
Collins, J. A., Shull, J. M., \& Giroux, M. L. 2004,
\apj, 605, 216

\bibitem[Danforth \& Shull (2005)]{DS05}
Danforth, C. W., \&  Shull, J. M. 2005,
\apj, 624, 555

\bibitem[Draine \& Anderson 1985]{Draine85}
Draine, B. D., \& Anderson, N. 1985, \apj, 292, 494

\bibitem[Gillmon \& Shull 2005]{GillShull05}
Gillmon, K., \& Shull, J. M. 2005, \apj, submitted 

\bibitem[Hartmann & Burton 1997]{HB97}
Hartmann, D., \& Burton, W. B. 1997, Atlas of Galactic Neutral Hydrogen,
(Cambridge: Cambridge Univ. Press)

\bibitem[Hollenbach et al. 1972]{Hollenbach71}
Hollenbach, D. J., Werner, M. W., \& Salpeter, E. E. 1971,
\apj, 163, 165

\bibitem[Hollenbach & McKee 1979]{Hollenbach79}
Hollenbach, D. J., \& McKee, C. F. 1979,
\apjs, 41, 555

\bibitem[Jura (1974)]{Jura74}
Jura, M.  1974, \apj, 191, 375

\bibitem[Lallement et al. 2003]{Lalle03}
Lallement, R.,  Welsh, B.Y., \etal\ 2003, ApJ, 411, 447

\bibitem[Lockman et al. (1986)]{Lock86}
Lockman, F. J., Jahoda, K., \& McCammon, D. 1986, ApJ, 302, 432

\bibitem[Lockman & Condon (2005)]{Lock05}
Lockman, F. J., \& Condon, J. J.  2005, AJ, 129, 1968

\bibitem[Low et al. (1984)]{Low84}
Low, F. J., \etal\ 1984, \apj, 278, L19

\bibitem[Moos et~al. (2000)]{Moos00}
Moos, H.~W., \etal\ 2000, \apjl, 538, L1

\bibitem[Penton et~al. (2000)]{Penton04}
Penton, S. V., Stocke, J. T., \& Shull, J. M. 2004, \apjs,
152, 29

\bibitem[Rachford (2002)]{R02}
Rachford, B., \etal\ 2002, \apj, 577, 221

\bibitem[Richter et al. (2001)]{Ri01}
Richter, P., Sembach, K. R., Wakker, B. P., \& Savage, B. D.
2001, \apj, 562, L181

\bibitem[Richter et al. (2003)]{Ri03}
Richter, P., Wakker, B. P., Savage, B. D., Sembach, K. R.
2003, \apj, 586, 230

\bibitem[Sahnow et al. (2000)]{sahnow00}
Sahnow, D. J., \etal\ 2000, \apjl, 538, L7

\bibitem[Savage, Bohlin, Drake, \& Budich (1977)]{S77}
Savage, B.~D., Bohlin, R.~C., Drake, J.~F., Budich, W. 1977,
\apj, 216, 291

\bibitem[Savage et al. (2002)]{Sav02}
Savage, B. D., Sembach, K. R., Tripp, T. M., \& Richter, P.
2002, \apj, 564, 631

\bibitem[Savage et al. (2003)]{Sav03}
Savage, B. D., \etal\ 2003, \apjs, 146, 125

\bibitem[Schlegel et al. (1998)]{Sch98}
Schlegel, D. J., Finkbeiner, D. P., \& Davis, M. 1998, ApJ, 500, 525

\bibitem[Sembach, K. R. et al. (2001b)]{Sem01a}
Sembach, K., Howk, J. C., Savage, B. D., Shull, J. M.,
Oegerle, W. D. 2001a, \apj, 561, 573

\bibitem[Sembach, K. R. et al. (2001b)]{Sem01b}
Sembach, K., Howk, J. C., Savage, B. D., Shull, J. M. 2001b,
\aj, 121, 992

\bibitem[Sembach, K. R. et al. (2003)]{Sem03}
Sembach, K. R., \etal\  2003, \apjs, 146, 165

\bibitem[Sembach, K. R. et al. (2003)]{Sem03}
Sembach, K. R., \etal\ 2003, \apjs, 150, 287

\bibitem[Shull, et al. (2000)]{Sh00}
Shull, J. M., \etal\ 2000, \apj, 538, L73

\bibitem[Shull, et al. (2004)]{S04}
Shull, J. M., Anderson, K. L., Tumlinson, J., \etal\ 2005,
\apj, submitted

\bibitem[Shull, Beckwith (1982)]{SB82}
Shull, J. M., \& Beckwith, S. V. W. 1982, \araa, 20, 163

\bibitem[Shull, Woods (1986)]{Sh86}
Shull, J. M., \& Woods, D. T. 1985, \apj, 280, 465

\bibitem[Snow, et al. (2000)]{Snow00}
Snow, T. P., \etal\ 2000, \apj, 538, L65

\bibitem[Spitzer, Jenkins (1975)]{SJ75}
Spitzer, L., \& Jenkins, E. B. 1975, \araa, 13, 133

\bibitem[Tumlinson et al. (2002)]{T02}
Tumlinson, J., \etal\ 2002, \apj, 566, 857

\bibitem[Tumlinson et al. (2004)]{T04}
Tumlinson, J., Shull, J. M., Giroux, M. L., \& Stocke, J. T.
2005, \apj, 620, 95

\bibitem[van Dishoeck \& Black (1986)]{vD86}
van Dishoeck, E. F., \& Black, J. H. 1986, \apjs, 62, 109

\bibitem[Wakker \& Schwarz (1991)]{Wa91}
Wakker, B. P., \& Schwarz, U. J. 1991, \aap, 250, 484

\bibitem[Wakker, et al. (2001)]{Wa01}
Wakker, B. P., \etal\ 2001, \apjs, 136, 537

\bibitem[Wakker, et al. (2002)]{Wa02}
Wakker, B. P., Oosterloo, T. A., \& Putman, M. E. 2002, \aj, 123, 1953

\bibitem[Wakker, et al. (2003)]{Wa03}
Wakker, B. P., \etal\ 2003, \apjs, 146, 1

\bibitem[Weiland, et. al. (1986)]{Wei86}
Weiland, J. L., Blitz, L., Dwek, E., Hauser, M. H.,
Magnani, L., \& Rickard, L. J. 1986, \apj, 306, L101

\bibitem[Welsh et al. (2002)]{Wel02}
Welsh, B. Y., Rachford, B. L., \& Tumlinson, J. 2002,
\aap, 381, 566

\bibitem[Wolfire et al. (1995)]{Wolf95}
Wolfire, M. G., McKee, C. F., Hollenbach, D. J., \& Tielens, A. G. G. M. 
   1995, \apj, 463, 673



\end{thebibliography}
\end{document}